\documentclass[fleqn,usenatbib]{mnras}

\usepackage{newtxtext,newtxmath}

\usepackage[T1]{fontenc}

\usepackage[normalem]{ulem}  
\newcommand\redout{\bgroup\markoverwith
{\textcolor{red}{\rule[0.5ex]{2pt}{0.8pt}}}\ULon}

\usepackage{tikz,xcolor,hyperref}

\definecolor{lime}{HTML}{A6CE39}
\DeclareRobustCommand{\orcidicon}{
	\begin{tikzpicture}
	\draw[lime, fill=lime] (0,0) 
	circle [radius=0.16] 
	node[white] {{\fontfamily{qag}\selectfont \tiny ID}};
	\draw[white, fill=white] (-0.0625,0.095) 
	circle [radius=0.007];
	\end{tikzpicture}
	\hspace{-2mm}
}

\foreach \x in {A, ..., Z}{\expandafter\xdef\csname orcid\x\endcsname{\noexpand\href{https://orcid.org/\csname orcidauthor\x\endcsname}
			{\noexpand\orcidicon}}
}

\usepackage{graphicx}	
\usepackage{amsmath}	
\usepackage{pdflscape} 
\usepackage{slantsc}

\title[Ionized nebulae around accreting WDs]{Linking the properties of accreting white dwarfs with the ionization state of their ambient medium}

\author[D. Souropanis et al.]{
D. Souropanis,$^{1,2,3\orcidA{}}$\thanks{E-mail: d.souropanis@noa.gr}
A. Chiotellis,$^{1\orcidB{}}$
P. Boumis,$^{1\orcidC{}}$
M. Chatzikos,$^{4\orcidD{}}$ 
S. Akras,$^{1\orcidE{}}$ 
L. Piersanti,$^{5,6\orcidI{}}$
\newauthor A. J. Ruiter,$^{7\orcidG{}}$
G. J. Ferland$^{4\orcidH{}}$\\
\\
$^{1}$Institute for Astronomy, Astrophysics, Space Applications
and Remote Sensing, National Observatory of Athens,
15236 Penteli, Greece\\
$^{2}$Department of Physics, National and Kapodistrian University of Athens, Panepistimiopolis, 15784 Zografos, Greece\\
$^{3}$ Isaac Newton Group of Telescopes, Apartado 321, E-38700 Santa Cruz de La Palma, Canary Islands, Spain\\
$^{4}$Department of Physics, University of Kentucky, Lexington KY 40506, USA\\
$^{5}$INAF - Osservatorio Astronomico d'Abruzzo, via Mentore Maggini, snc, I-64100 Teramo, Italy\\  
$^{6}$INFN - Sezione di Perugia, Via A. Pascoli snc, I-06123 Perugia, Italy\\
$^{7}$School of Science, University of New South Wales,
Australian Defence Force Academy, Canberra, ACT 2600, Australia\\
}

\date{Accepted XXX. Received YYY; in original form ZZZ}

\pubyear{2022}

\begin{document}
\label{firstpage}
\pagerange{\pageref{firstpage}--\pageref{lastpage}}
\maketitle

\begin{abstract}

Steadily accreting white dwarfs (WDs) are efficient sources of ionization and thus, are able to create extended ionized nebulae in their vicinity. These nebulae represent ideal tools for the detection of accreting WDs, given that in most cases the source itself is faint. In this work, we combine radiation transfer simulations with known H and He accreting WD models, providing for the first time the ionization state and the emission line spectra of the formed nebulae  as a function of the WD mass, the accretion rate and the chemical composition of the accreted material. We find that the nebular optical line fluxes and radial extent  vary strongly with the WD’s accretion properties, peaking in systems with WD masses of~0.8 -- 1.2~$\rm~M_{\odot}$. Projecting our results on the ‘BPT' diagnostic diagrams, we show that accreting WDs nebulae  possess characteristics distinct from those of \ion{H}{II}-like regions, while they share similar line ratios with the galactic low-ionization emission-line regions. Finally, we compare our results to the relevant constraints imposed by the lack of ionized nebulae in the vicinity of supersoft X-ray sources (SSSs) and Type Ia supernova remnants -- sources  which are related to
steadily-accreting WDs.  The large discrepancies uncovered by our comparison rule out any steadily-accreting WD as a potential progenitor of the studied remnants and additionally require the ambient medium around the SSSs to  be less dense than 0.2~$\rm~cm^{-3}$.  We discuss possible alternatives that could bridge the incompatibility between the theoretical expectations and the relevant observations.

\end{abstract}

\begin{keywords}

(stars :) white dwarfs --  binaries: general  -- ISM: supernova remnants -- radiation mechanisms: general -- line: formation

\end{keywords}



\section{Introduction}\label{sec:intro}

White Dwarfs (WDs) are the compact degenerate remnants of low- and intermediate-mass stars ($ \sim 1-8~\rm M_{\odot}$), representing the endpoints of stellar evolution for the vast majority of stars in the Universe. Having ceased any nuclear reaction in their center, isolated WDs are doomed to  quiescently fade out by radiating their internal energy. Nevertheless, a high percentage of WDs are members of interacting binary systems \citep[e.g.][]{Toonen2017,Holberg2009}. In these systems hydrogen or helium rich material can be transferred from the companion star to the WD, altering entirely its properties and expected evolution. Depending on the accretion process, the chemical composition of the accreting material, the WD mass and the nature of the companion star, these binary systems can reveal several intriguing phenomena and up to date they have been linked to a  number of astrophysical objects such as classical, recurrent and He novae  \citep[e.g.][]{Gallagher,Kato, 2021kemp},  supersoft X-ray sources \citep[][]{VandenHeuvel1992, Nomoto2007}, symbiotic binaries \citep[e.g.][]{ Mikolajewska2007, Mohamed2012,  Akras2019} and  AM CVn systems \citep[e.g.][]{pazy, Nelemans2001, pier15}. 

Carbon-Oxygen WDs (CO WDs) in interacting binaries have also been proposed to be the progenitors of the cosmic explosions known as Thermonuclear or Type Ia Supernovae (SNe~Ia).  It has been suggested that the explosion of the WD results either through mass accretion from a non degenerate companion star \citep[{\it single degenerate scenario},][]{Whelan1973} or from the merger with another WD \citep[{\it double degenerate scenario},][]{Iben1984}.  While each scenario may explode through one or more explosion mechanism, it is not clear so far which explosion mechanism is responsible (or dominant) in producing SNe~Ia \citep[see][for a recent review]{Ruiter2020}. 
In the `Chandrasekhar-mass scenario', the required central density capable of initiating the carbon ignition is reached when the WD approaches the Chandrasekhar mass limit ($ M_{\rm WD}\sim~1.38~\rm M_{\odot}$). In an alternative scenario, the double-detonation 
scenario, an initial detonation of a He layer on a sub-Chandrasekhar mass WD triggers a shock wave that rapidly propagates inside the WD's interior and compresses the central carbon, facilitating the onset of explosive carbon burning \citep[e.g.][]{Livne1990, Fink2010, Sim2010,Woosley2011, Hillebrandt2013}. Simulations have shown that the physical conditions adequate to reproduce `normal' SNe~Ia via the double-detonation mechanism are very favorable for CO WDs in the mass range $\approx ~0.9 -1.0~\rm M_{\odot}$ \citep[see][for details]{Piro2014,Shen2018}.

Given that CO WDs are typically formed with lower masses  \citep[the mass distribution of single CO WDs is peaked at $M~\sim~0.56~\rm M_{\odot}$,][]{Madejj2004}, this implies that -at least for the vast majority of SNe~Ia progenitors- mass accretion is required in order the WD to reach the appropriate mass for the explosion.
For such progenitors, it has been shown that if the accretion rate is on the order of $\rm \sim a~few~\times 10^{-7} ~\rm M_{\odot}~yr^{-1}$  and $\rm \sim a~few~\times 10^{-6}~\rm M_{\odot}~yr^{-1}$   for WDs accreting hydrogen- or helium- rich material, respectively, the shell nuclear burning occurs steadily and the WD will increase in mass  \citep[e.g.][]{cassisi1998,shen2007,Nomoto2007,wolf2013,PIERSANTI2014}. 
WDs in such an accretion mode are characterized by surface temperatures exceeding $10^5$~K.  This means that a significant flux of UV and soft X-ray thermal photons are expected to be radiated from the surface of the accreting WD. In turn, these photons will photo-ionize the ambient medium in the vicinity, shaping extended ionized nebulae bright in the \ion{He}{II} 4686 \AA, [\ion{O}{III}] 5007 \AA, and [\ion{O}{I}]~6300~\AA  \ emission lines \citep{Rappaport1994}. These nebulae should persist long after  \citep[$t> 10^4$~yrs, ][]{kuuttila2019, Woods2018} the WD accretion phase has ceased, until the majority of the ionized gas has recombined. For higher WD masses  ($> 1.1~\rm M_{\odot}$), the surface temperature can reach up to a few$~\times~10^{6}$~K, and hence, the WD’s photosphere becomes bright in the soft X-ray band \citep[0.3 – 0.7 keV;][]{VandenHeuvel1992, Rappaport1994a}. In fact several permanent supersoft X-ray sources (SSSs) have been identified, for many of which the optical and X-ray follow up observations revealed that they are binary systems containing a WD that steadily burns the accreted material transferred by its companion star \citep{VandenHeuvel1992, Rappaport1994a, Kahabka1997}. Given that SSSs consist of a massive steadily-accreting WD, these systems have been proposed as a potential prominent scenario for SNe Ia progenitors \citep{Hachisu1999, Li1997}.

If indeed the WDs in progenitors of SNe Ia have passed through a steady accretion phase prior to their final explosion then SNe Ia and their young remnants ($t_{\rm SNR} \le 10^{4} \ \rm years$; hereafter SNRs Ia) should be surrounded by extended optically bright nebulae. Based on this fact, several independent studies were conducted in parallel aiming to observe ionized nebulae around individual nearby SNe~Ia \citep{2019Graur} and young SNRs~Ia \citep[e.g.][]{kuuttila2019,Farias2020}. Nevertheless, none of them detected any nebular optical emission around the studied objects and thus, strict limits were placed on the ionizing flux of their WD progenitors. An additional constraint of the ionizing flux of the SN Ia progenitors results by the presence of Balmer dominated shocks  that most SNRs Ia display.   Given that the Balmer lines indicate that the SNR is evolving into a --at least partially-- neutral ambient medium, hot luminous WD progenitors seem incapable of explaining the properties of the vast majority of SNRs Ia \citep{woods2017,Woods2018}. 

Even more surprising is that, with the exception of CAL 83, ionized emission nebulae have not been detected around other SSSs  \citep{remillard1995, Farias2020}. The lack of any nebular optical emission around the studied SSSs has been attributed either to a very low density ambient medium  or to an extremely transient SSS phase \citep{Farias2020, 2016Woods, Rappaport1994}. However, to date  there  is little understanding of both the evolution of these systems and the properties of their surrounding interstellar medium . Regarding the observed nebula around CAL~83, its origin remains debatable as current models reveal difficulties in reproducing its optical emission line ratios \citep{Gruyters2012}. A better understanding on the processes that govern the interaction of accreting WDs with their surrounding gas is an important task given that in most cases the source itself is heavily obscured \citep[][]{Chen2015}.

Past efforts in modeling the ionization state around accreting WDs  focused on the class of SSSs. Thereby,  these models were limited to the most massive and hottest members of the  accreting WD family.  In addition, for the description of the ionizing source, random pairs of luminosities and effective temperatures were adopted within the expected range that accreting WDs display \citep[]{Rappaport1994}. Nevertheless, detailed models of steady nuclear-burning accreting WDs show that the WD's luminosity and  effective temperature are strongly correlated with each other, and in turn strongly depend on the WD mass and mass accretion rate \citep[e.g.][]{Nomoto2007, PIERSANTI2014}.

Aiming to further investigate the interaction of accreting WDs with their ambient medium and to extend models from previous works, in this paper we link the properties of H and He accreting WDs with those of the surrounding ionized nebulae by coupling known WD accretion models from the literature with photo-ionization numerical tools. We explore the whole parameter space of steadily-accreting WDs and we present luminosities and surface brightness profiles of various important nebular emission lines formed around these sources. In addition, we extend our study to the case of He-accreting WDs, systems that have not been studied in the past. In this way, we provide observational  predictions for various nebular emission lines as a function of the WD mass, the accretion rate and the chemical composition of the accreted material.   In the second part of this work, we compare  our results with the observational constraints that exist as the result of lack of ionized nebulae found around SSSs and SNRs~Ia, (re-)assessing the compatibility of steadily nuclear-burning WDs  as plausible progenitors for these objects. 

This paper is organized as follows.  In Section \ref{sec:2}  we describe our radiation transfer simulations and we present the results of our modelling regarding the ionization state and spectral properties of ionized nebulae around steadily-accreting WDs. In Section \ref{sec:3} we compare the results of our modeling to the relevant observations from SSSs and SNRs Ia. We discuss and summarise our main results and conclusions in Sections \ref{sec:4} and \ref{sec:5}.

\section{Modelling the ionized nebulae around steadily-accreting White Dwarfs}\label{sec:2}

\begin{figure}
\includegraphics[trim=0cm 0cm 0cm 0cm, clip=true,width=\columnwidth,angle=0]{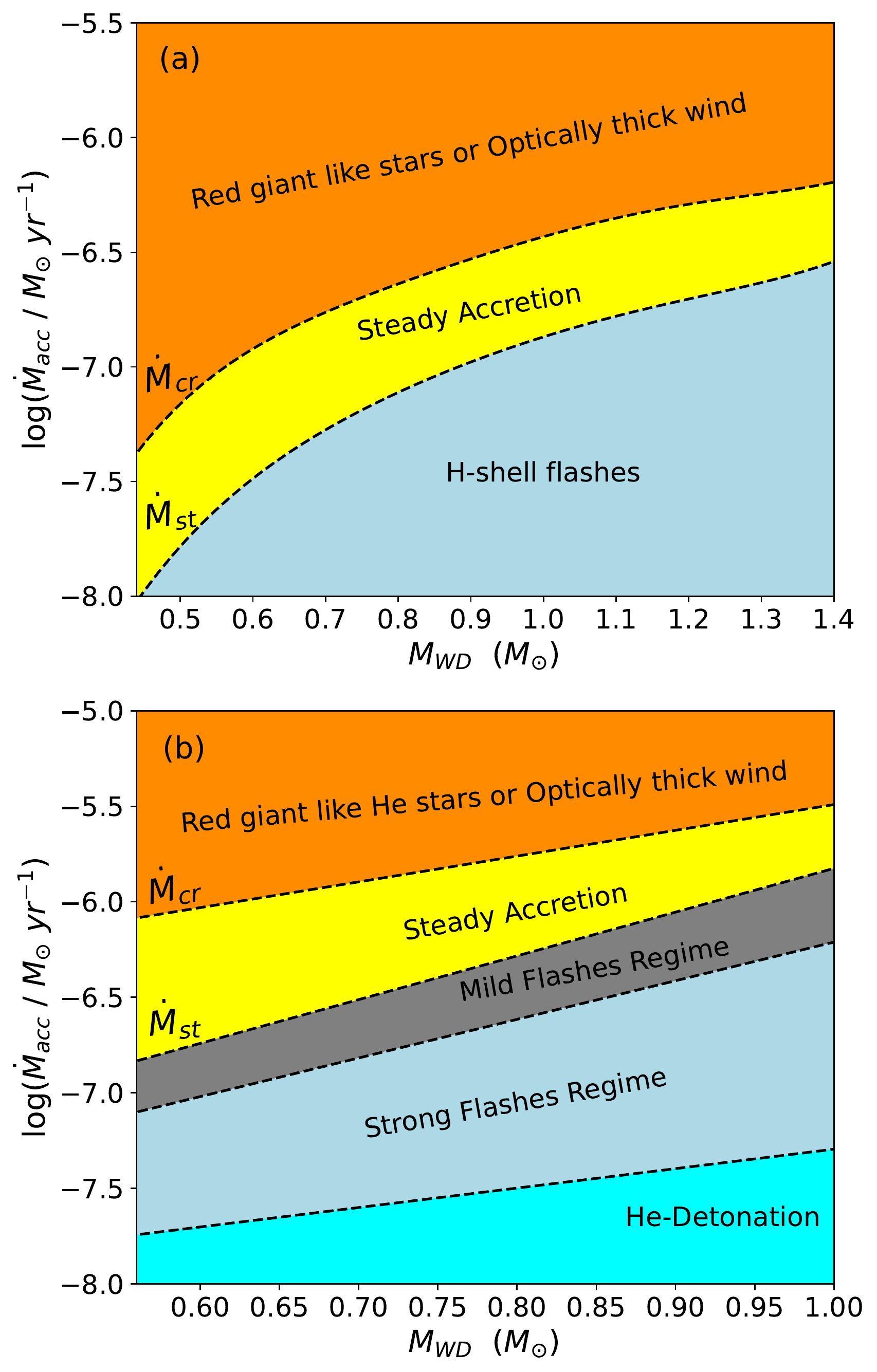} 
\caption { The possible accretion regimes for WDs accreting H (a) and He (b) extracted by the models of \citet{Nomoto2007} and \citet{PIERSANTI2014}, respectively. The dashed lines represent the  borders between the regimes.
}
\label{fig:regimes}
\end{figure}

\subsection{Methodology}\label{subsec:2.1}

To model the interaction of accreting WDs with their ambient medium, we use the photoionization code {\sc cloudy} \citep[v17.02, ][]{Ferland2017}. {\sc cloudy} is an open-source spectral synthesis code designed to simulate conditions in interstellar matter under a broad range of conditions (i.e. gas density, composition, incident spectrum etc). More details, source files and all necessary data are available from \url{www.nublado.org}.

To link the WD’s mass accretion properties with those of the ionized ambient medium, we use the results produced from the accreting WD models of \citet{Nomoto2007} and \citet{PIERSANTI2014}. \citet{Nomoto2007} studied the properties of CO WDs accreting H rich material in and near the steady accretion regime and they constructed steady models for various accretion rates and WD masses ranging from $0.5$~to~$1.38~\rm M_{\odot}$. \citet{PIERSANTI2014} studied the long term evolution of He accreting CO WDs for initial WD masses and accretion rates in the range $0.60-1.02~\rm M_{\odot}$ and $10^{-9}-10^{-5}~ \rm M_{\odot}yr^{-1}$, respectively. We refer to \citet{Nomoto2007} and \citet{PIERSANTI2014} for the full details of their models, while Figure \ref{fig:regimes} shows the possible accretion regimes as a function of the WD  mass ($M_{\rm WD}$) and the accretion rate ($\Dot{M}_{\rm acc}$), as extracted by their works for H and He accreting WDs, respectively.

The steady accretion regime  - the regime explored in this study-  is confined in a narrow strip between the so-called stable and critical accretion rates ($\dot{M}_{\rm st}$ and $\dot{M}_{\rm cr}$).  In this regime the mass transfer is conservative as the rate at which H/He is converted via surface nuclear burning into He/CO-mixture is equal or at least fairly close to  the accretion rate.   Consequently, the WD's intrinsic luminosity ($L_{\rm WD}$) and its effective temperature ($T_{\rm eff}$) are strongly paired to each other and dominantly determined from  the mass-accretion rate and the WD mass. 
The results extracted by the aforementioned accreting WD models, namely the pairs of values for the WD’s intrinsic luminosity and effective temperature as a function of the WD mass and accretion rate, are used as input parameters for the ionizing source in {\sc cloudy} (see Appendix A). The central source is assumed to be a blackbody, which provides a reasonable approximation of the ionizing emission of nuclear-burning WDs, except far into the Wien tail \citep[][]{woods2017,Chen2015}. We further assume a spherical symmetry in our models, a constant density ambient medium and solar gas abundances \citep[see the default solar values as defined in {\sc cloudy}, which are taken from][]{Grevesse1998, AllendePrieto, holweger2001}. The ambient medium number density in our models is taken to be $0.2~\rm cm^{-3}$ and $ 2~\rm cm^{-3}$, 
which are plausible values to use in the vicinity of SSSs \citep[e.g.][]{Farias2020} and of  young SNRs~Ia \citep[e.g.][]{Yamaguchi2014,martinez2018}. 
We terminate our calculations when the gas temperature drops below 3000 K (at these low nebular temperatures the hydrogen ionization fraction falls below 10\%, and thus its contribution to the total nebular line emission is negligible).

Given these assumptions, we estimate the predicted surface brightness profile and luminosity in any emission line produced by the surrounding gas, for any given pair of the WD mass and accretion rate  within the stable accretion regime. In particular, {\sc cloudy} provides the volume emissivity $e(r)$ of any emission line as a function of the distance $r$ from the ionizing source. The surface brightness in a nebular spectral line, denoted by $i$, depends on the line of sight  integral of the emission coefficient, which is given by:
  \begin{equation}
      SB_{i}(r)=  \int_{s} \frac{\epsilon_{i}(r)}{4\pi} \,ds,
  \end{equation}
  where the integral is taken along the line of sight through the nebula at impact parameter  \textit{r} from the centre of the nebula. The predicted total line luminosity  is given by
integrating the surface brightness over the area of the nebula, which is equivalent to integrating the emission coefficient over the volume of the ionized nebula, which is given by:
\begin{equation}
    L_{i}=\int_{0}^{r} SB_{i}(r) 2\pi rdr .
\end{equation}

\begin{landscape}

\begin{table}

\centering

\caption{Predicted nebular radii and luminosities of prominent optical emission lines surrounding H accreting WDs as a function of WD mass for the two stability boundaries  of accretion rate  in the steady accretion regime based on models of \citet{Nomoto2007}, and embedded in constant ISM densities of 0.2 and 2 $\rm cm^{-3}$ .}
\label{tab:ModelA}
\begin{tabular}{|c|c|c|c|c|c|c|c|c|c|c|c|c|c|c|}

\multicolumn{15}{|c|}{H accreting WDs}                  \\ \hline
 M    & $\Dot{M}_{\rm acc}$    & log$(L/L_{\odot})$ & log$(T_{\rm eff}/\rm K)$ & $R_{H^{0}, 0.5}$ & $R_{He^{0}, 0.5}$ &   $\ion{He}{II}$ &  $[\ion{O}{III}]$ &$[\ion{O}{I}]$& $[\ion{O}{II}]$ &H$\alpha$ & H$\beta$& $[\ion{N}{II}]$ &  $[\ion{N}{II}]$& $[\ion{S}{II}]$ \\ 
  & $( \times~10^{-7}$ & & &(pc) & (pc) &  4686 {\AA} & 5007 {\AA} & 6300 {\AA} & 3727 {\AA} &  6562 {\AA} &  4861 {\AA} & 5754 {\AA} & 6584  {\AA} & 6717 {\AA}   \\
  ($\rm M_{\odot}$) &  $\rm M_{\odot}~yr^{-1}$)  & & &  & &$(\rm erg\ s^{-1})$ &  $\rm (erg\ s^{-1})$ & $\rm (erg\ s^{-1})$ & $\rm (erg\ s^{-1})$ & $\rm (erg\ s^{-1})$ & $\rm (erg\ s^{-1})$ & $\rm (erg\ s^{-1})$ & $\rm (erg\ s^{-1})$ & $\rm (erg\ s^{-1})$  
 \\ \hline
 \multicolumn{15}{|c|}{$n_{\rm ism}$ = 2 $\rm cm^{-3}$}     \\ \hline
 0.6  & 0.3258 & 3.38  &  5.39 & 9.72
  & 11.7 & 2.03e+34 &   2.72e+35
    &  4.28e+34 &  4.72e+35 & 1.45e+35 &5.15e+34 & 2.76e+33  & 1.85e+35 & 1.69e+35 \\ 
 0.6  & 1.2    & 3.95  &  4.87  & 16.3 & 17.3
 & 7.26e+33 &   5.92e+35   & 3.29e+34 & 7.10e+35  & 9.27e+35 & 3.30e+35 & 2.73e+33 &  3.85e+35 & 2.89e+35\\ 
 0.8  & 0.775  & 3.76  &  5.57  & 11.3 & 14.3 & 3.85e+34 &   4.83e+35    &  1.17e+35 & 9.69e+35 &  2.62e+35 & 9.18e+34 & 6.17e+33 & 3.69e+35 & 2.93e+35 \\ 
 0.8  & 2.3    & 4.23  &  5.44  &18.4
  & 21.2 &  1.59e+35  &  2.78e+36    &  2.61e+35 & 3.17e+36 &  1.01e+36 & 3.57e+35 & 1.94e+34  & 1.16e+36 & 9.77e+35 \\ 
 1.0    & 1.35   & 4.00  &  5.72  & 11.7   & 15.8 & 4.51e+34 &   5.19e+35    & 2.06e+35 & 1.30e+36 &  3.49e+35 & 1.20e+35 & 8.71e+33 & 5.00e+35 & 3.06e+35 \\ 
 1.0    & 3.7    & 4.45  &  5.54 & 20
 & 24.2
 &  2.24e+35 &     3.75e+36   & 5.22e+35& 4.88e+36  &  1.41e+36 & 4.94e+35 & 3.14e+34  & 1.77e+36 & 1.28e+36 \\ 
 1.25 & 2.15   & 4.20  &  5.91  & 10.6 & 15.5
 & 3.56e+34 &   3.48e+35    &  2.70e+35& 1.25e+36 &   3.62e+35 & 1.22e+35 & 8.85e+33 & 5.10e+35 & 2.43e+35 \\ 
 1.25 & 5.4    & 4.64  &  5.67 & 21.1  & 26.9
 &2.71e+35 &   4.39e+36   &  9.08e+35 & 6.70e+36
&  1.86e+36 & 6.40e+35 & 4.49e+34 & 2.44e+36 & 1.40e+36 \\ 
 1.35 & 2.55   & 4.28   & 6.03 & 9.3 & 13.9  &  2.37e+34 &   2.04e+35    & 2.44e+35  & 9.47e+35 &  3.03e+35 & 1.01e+35 & 7.08e+33 & 4.22e+35 & 1.87e+35  \\ 
 1.35 & 6      & 4.70  &  5.81 & 18.8
  & 25.6 & 1.98e+35  &   2.84e+36   & 1.08e+36 & 6.10e+36 &  1.66e+36 & 5.61e+35 & 4.19e+34 & 2.29e+36 & 1.03e+36 \\ 
1.38 & 2.75   & 4.31  &  6.10 & 8.5  & 12.7 &  1.82e+34  &  1.46e+35    &  2.17e+35 & 7.75e+35 &   2.64e+35 & 8.79e+34 & 5.96e+33 & 3.62e+35 & 1.60e+35\\ 
 1.38 & 6.2    & 4.71  &  5.87 & 17.5  & 24.6 &  1.61e+35    &   2.17e+36   &  1.05e+36 &  5.28e+36 &  1.50e+36 & 5.03e+35 & 3.75e+34 & 2.08e+36 & 8.81e+35 \\ \hline
 \multicolumn{15}{|c|}{$n_{\rm ism}$ = 0.2 $\rm cm^{-3}$}     \\ \hline
 0.6  & 0.3258 & 3.38  &  5.39& 43 & 57.4 &  1.58e+34  & 1.25e+35    &  4.96e+34 & 4.05e+35 &  1.33e+35  & 4.72e+34 & 2.34e+33 & 1.79e+35 & 1.83e+35 \\ 
 0.6  &   1.2  & 3.95  &  4.87 & 76 & 83.1   &  7.06e+33  &   3.28e+35   &  3.68e+34 & 6.17e+35 &  9.27e+35  & 3.28e+35 & 2.22e+33 &  4.05e+35 & 3.36e+35 \\ 
 0.8  &  0.775 & 3.76  &  5.57  & 49.3 & 67.9 &  2.91e+34  &   2.10e+35   & 1.18e+35 & 7.94e+35 &  2.26e+35 & 7.92e+34 & 4.95e+33 & 3.33e+35 & 3.12e+35  \\ 
 0.8  &   2.3  & 4.23  &  5.44 & 82.3 & 103.4  &   1.31e+35 &    1.52e+36   &  3.32e+35 & 3.17e+36 &   9.29e+35  & 3.29e+35 & 1.90e+34 & 1.27e+36 & 1.17e+36 \\ 
 1.0    &1.35    & 4.00  &  5.72 & 49.9   & 71.6
 &  3.29e+34 &  2.10e+35    &  1.75e+35& 9.77e+35 &  2.79e+35 & 9.66e+34 & 6.38e+33 & 4.07e+35 & 3.17e+35 \\ 
 1.0    &  3.7   & 4.45  &  5.54 & 88.9 &114.8    & 1.79e+35 &  1.95e+36    &  5.81e+35 & 4.63e+36 & 1.26e+36 & 4.41e+35 &  2.92e+34  & 1.81e+36 & 1.52e+36  \\ 
 1.25 &2.15    & 4.20  &  5.91 & 43.3 & 67  &  2.41e+34 &   1.24e+35    &  1.86e+35 & 8.07e+35 &  2.55e+35 & 8.69e+34 & 5.56e+33 & 3.51e+35 & 2.19e+35 \\ 
 1.25 &5.4     & 4.64  &  5.67 & 91.9 & 124.8
 &  2.11e+35 &   2.16e+36   & 8.87e+35 & 5.95e+36 &  1.59e+36 & 5.51e+35 & 3.88e+34 & 2.32e+36 & 1.65e+36    \\ 
 1.35 &2.55    & 4.28  &  6.03 & 36.7 & 57.9  &   1.51e+34&     6.59e+34   &  1.48e+35 & 5.54e+35 &   1.92e+35 & 6.50e+34 & 3.98e+33 & 2.55e+35  & 1.46e+35 \\ 
 1.35 & 6      & 4.70  &  5.81 & 80  & 115  &  1.46e+35   &  1.24e+36    &  8.81e+35 & 4.65e+36 &   1.31e+36& 4.47e+35 & 3.16e+34 & 1.88e+36 & 1.11e+36    \\ 
 1.38 &2.75    & 4.31  &  6.10 & 33.5  & 53.1  &   1.12e+34&   4.51e+34   &  1.23e+35 & 4.29e+35 &  1.56e+35 & 5.26e+34 & 3.17e+33 & 2.06e+35 & 1.15e+35 \\ 
 1.38 & 6.2    & 4.71  &  5.87 & 73.7  & 107.4 & 1.16e+35&   8.93e+35    &  8.04e+35 & 3.87e+36 &  1.13e+36 & 3.83e+35 & 2.67e+34 & 1.60e+36 & 8.89e+35   \\ \hline
\label{tab:hydrogen}
\end{tabular}
\end{table}
\end{landscape}

\begin{landscape}
\begin{table}

    \caption{Predicted nebular radii and luminosities of prominent optical emission lines surrounding He accreting WDs, as a function of WD mass for the two stability boundaries  of accretion rate   in the steady accretion regime based on models of \citet{PIERSANTI2014}, and embedded in constant ISM densities of 0.2 and 2 $\rm cm^{-3}$ .} 
    
\centering

\begin{tabular}{|c|c|c|c|c|c|c|c|c|c|c|c|c|c|c|}

\multicolumn{15}{|c|}{He accreting WDs}                 \\ \hline
 M    & $\Dot{M}_{\rm acc}$   & log$(L/L_{\odot})$ & log$(T_{\rm eff}/\rm K)$& $R_{H^{0}, 0.5}$ & $R_{He^{0}, 0.5}$ &$\ion{He}{II}$ & $[\ion{O}{III}]$ &$[\ion{O}{I}]$& $[\ion{O}{II}]$ &H$\alpha$ & H$\beta$& $[\ion{N}{II}]$ & $[\ion{N}{II}]$& $[\ion{S}{II}]$ \\ 
  & $( \times~10^{-6}$ & & &(pc) & (pc) & 4686 {\AA} & 5007 {\AA} & 6300 {\AA} & 3727 {\AA} &  6562 {\AA} &  4861 {\AA} & 5754 {\AA} & 6584  {\AA} & 6717 {\AA}  \\
  ($\rm M_{\odot}$) &  $\rm M_{\odot}~yr^{-1}$)  & & && & $(\rm erg\ s^{-1})$ & $\rm (erg\ s^{-1})$ & $\rm (erg\ s^{-1})$ & $\rm (erg\ s^{-1})$ & $\rm (erg\ s^{-1})$ & $\rm (erg\ s^{-1})$ & $\rm (erg\ s^{-1})$ & $\rm (erg\ s^{-1})$ & $\rm (erg\ s^{-1})$  
 \\ \hline
\multicolumn{15}{|c|}{$n_{\rm ism}$ = 2 $\rm cm^{-3}$}     \\ \hline
0.6 & 0.15 & 3.33  &  5.28   &9.9 &11.5
   & 1.83e+34   & 2.77e+35    &  3.21e+34  & 4.14e+35 & 1.52e+35 & 5.42e+34 & 2.31e+33 & 1.70e+35 & 1.52e+35  \\ 
 0.6 & 0.95 & 4.03  &  4.78  & 16.7& 17.3 & 2.37e+33 &  3.85e+35   &  1.82e+34  & 5.15e+35 & 1.10e+36 & 3.89e+35 & 1.65e+33 & 3.29e+35  & 2.34e+35 \\ 
 0.7 & 0.3  & 3.73  &  5.43  & 12.4 & 14.9   &  4.70e+34&   6.92e+35   &  9.44e+34  & 1.04e+36 & 3.13e+35 & 1.11e+35 & 6.29e+33 &  3.97e+35 & 3.49e+35 \\ 
 0.7 & 1.80 & 4.34  &  4.80 &21.4 & 22.2  & 7.28e+33 &  1.13e+36   &  4.52e+34  & 1.25e+36 & 2.25e+36 & 7.98e+35 & 4.34e+33 & 7.11e+35 & 4.99e+35 \\ 
 0.8 & 0.52 & 4.03  &  5.56 &14.3 &17.8  &  7.91e+34 &  1.13e+36    &  2.14e+35  & 1.88e+36 & 5.20e+35 & 1.82e+35 & 1.20e+34 & 7.01e+35 & 5.41e+35 \\ 
 0.8 & 1.90 & 4.49  &  5.05 & 25.6& 27.1  &   1.45e+35 &  5.62e+36    &  2.10e+35  & 4.29e+36 & 2.96e+36 & 1.06e+36 & 2.12e+34 &  1.79e+36 & 1.39e+36 \\ 
  0.9 & 0.71 & 4.26  &  5.69 & 15.1 & 19.9 &  9.73e+34 &  1.34e+36    &  3.84e+35  & 2.65e+36 & 7.10e+35 & 2.45e+35 & 1.77e+34 & 9.93e+35 & 6.01e+35\\ 
  0.9 & 2.70 & 4.63  & 5.25  & 27.6 & 29.4&  4.10e+35 &   9.93e+36   &  3.54e+35  & 7.00e+36 & 3.30e+36 & 1.19e+36 & 3.92e+34 &  2.61e+36 & 2.12e+36 \\ 
1.0 & 1.40 & 4.40  & 5.77  & 15.3 & 20.8 & 1.04e+35  &  1.37e+36   &  5.26e+35  & 3.10e+36 & 8.45e+35 & 2.88e+35 & 2.13e+34 & 1.18e+36 & 6.07e+35  \\ 
 1.0   & 2.90 & 4.71  & 5.44 & 27 & 30.1  &   5.21e+35  &   1.04e+37   &  6.74e+35  & 9.34e+36 & 3.14e+36 & 1.11e+36 & 5.58e+34 & 3.19e+36 & 2.59e+36 \\  \hline
 \multicolumn{15}{|c|}{$n_{\rm ism}$ = 0.2 $\rm cm^{-3}$}     \\ \hline
 0.6 & 0.15 & 3.33  & 5.28 & 44.4& 56.9
  & 1.46e+34  &   1.36e+35   &  3.98e+34  & 3.73e+35 & 1.43e+35 & 5.13e+34 & 2.03e+33 & 1.72e+35 & 1.71e+35  \\ 
 0.6 & 0.95 & 4.03  & 4.78 & 77.8 &	81.9
  & 2.33e+33 &   2.08e+35   &  1.99e+34  & 4.17e+35 & 1.10e+36 & 3.88e+35 & 1.25e+33 & 3.31e+35 & 2.65e+35  \\ 
 0.7 & 0.3  & 3.73  & 5.43 & 55.6 &	72.6
  &  3.71e+34  &   3.36e+35   &  1.12e+35  & 9.46e+35 & 2.86e+35 & 1.014e+35 & 5.61e+33  & 3.99e+35 & 3.91e+35  \\ 
 0.7 & 1.80 & 4.34  & 4.80 & 99.6 &	105.3
   & 7.23e+33 &   6.41e+35   &  5.15e+34  & 1.10e+36 & 2.25e+36 & 7.96e+35 & 3.59e+33 & 7.69e+35 & 5.96e+35  \\ 
 0.8 & 0.52 & 4.03  & 5.56  & 62.6 &	84.5  &  6.14e+34 &   5.33e+35   &  2.25e+35  & 1.65e+36 & 4.55e+35 & 1.60e+35 & 1.03e+34 & 6.70e+35 & 6.03e+35  \\ 
 0.8 & 1.90 & 4.49  & 5.05  & 118.6 &	131.7 & 1.40e+35 &   3.66e+36   &  2.83e+35  & 4.54e+36 & 2.94e+36 & 1.05e+36 & 2.16e+34 & 2.14e+36 & 1.78e+36 \\ 
 0.9 & 0.71 & 4.26  & 5.69 & 64.9 &	91.7  & 7.41e+34  &   5.93e+35   &  3.51e+35  & 2.15e+36 & 5.90e+35 & 2.05e+35 & 1.40e+34 & 8.71e+35 & 6.61e+35  \\ 
 0.9 & 2.70 &  4.63 & 5.25   & 127.1 &	143.36
& 3.71e+35 &  6.65e+36    &  5.41e+35  & 8.01e+36 & 3.23e+36 & 1.16e+36 & 4.41e+34 & 3.27e+36 & 2.82e+36 \\ 
 1.0   & 1.40 & 4.40  & 5.77  &  65.2&	92.9
& 7.69e+34 &   5.78e+35   &  4.37e+35  & 2.38e+36 & 6.68e+35 & 2.30e+35 & 1.59e+34 & 9.70e+35 & 6.44e+35  \\ 
 1.0   & 2.90 & 4.71  &  5.44  & 121 &	145.5 & 4.41e+35 &   6.34e+36   &  9.20e+35  & 9.88e+36 & 2.94e+36 & 1.04e+36 & 5.98e+34 & 3.78e+36 & 3.31e+36 \\  \hline
\label{tab:helium}
\end{tabular}
\end{table}
\end{landscape}

\subsection{Results}\label{subsec:2.2}

In order to offer direct links between the properties of H and He accreting WDs with those of their ambient medium, we present in Tables~\ref{tab:hydrogen}~and~\ref{tab:helium}, respectively, the  luminosities of the most important optical diagnostic nebular emission lines as a function of WD accretion properties (i.e the WD's mass, accretion rate, intrinsic luminosity and effective temperature) and for the two studied interstellar medium (ISM) densities ($n_{\rm ism}= 0.2$ and 2~$\rm~cm^{-3}$). As an index for the ionizing efficiency of  steadily-accreting, nuclear burning WDs, for each combination of WD mass and accretion rate we also provide the parameters  $R_{\rm H^{0},~0.5}$ and $R_{\rm He^{0},~0.5}$ (fifth and sixth columns of Tables~\ref{tab:hydrogen}~and~\ref{tab:helium}), which indicate the radial distance from the central source where the neutral H and He  are equal to the corresponding ionized component(s) i.e. 50 \% each. Given the large parameter space involved in the  results tabulated in Tables  \ref{tab:hydrogen} and \ref{tab:helium}, the variability  of the nebular emission lines' luminosity and the radial extent of the regions of ionized H and ionized He as a function of the studied physical variables can not be firmly drawn. However, the following distinctive patterns can be individuated: 
\\
{\bf i)} On average, the brightest and most prominent optical emission lines of the ionized nebulae  of the studied H/He nuclear burning  WD models - in  descending  order - are:    $[\ion{O}{ii}]$ 3727~\AA ~\footnote{The line $[\ion{O}{ii}]$ 3727~\AA \ refers to the combined emission of  $[\ion{O}{ii}]$~3726.03~\AA \ and $[\ion{O}{ii}]$~3728.73 \AA.}, $[\ion{O}{iii}]$~5007~\AA, $\rm H_{\alpha}$, $[\ion{N}{ii}]$ 6584 \AA \   and $[\ion{S}{ii}]$ 6717 \AA \ . Concerning the radial extent of the spheres of ionized H and He around accreting WDs, we find that the quantity $R_{\rm He^{0},~0.5}$ is about 1.1 -1.6 larger than the $R_{\rm H^{0},~0.5}$ in all of our models  at both tested densities. 
\\
{\bf ii)} The increase of the WD's accretion rate from its minimum value ($\dot{M}_{\rm st}$) to its maximum ($\dot{M}_{\rm cr}$)  leads to a substantial enhancement for almost all line luminosities by about one order of magnitude. Exceptions to this trend are the lines of \ion{He}{II}~4686 \AA \ , $[\ion{O}{i}]$~6300~ \AA\ and  $[\ion{N}{ii}]$ 5754 \AA \ for which the increase of the mass accretion rate is accompanied by a modest decrease of their luminosity for the case of low mass   H- and He-accreting WDs ($M_{\rm WD} \lesssim 0.8~\rm M_{\odot}$). The principal impact on the sizes of H and He ionized regions around accreting WDs is the formation of  more extended nebulae, whose radii are predicted to increase by a factor of two when the accretion rate increases from its minimum value to the maximum one.
\\
{\bf iii)} Apart from the accretion rate, the  line luminosities are highly sensitive to the WD mass.  The brightest and most extended ionized nebulae around H-accreting WDs are expected to accompany systems with WD masses ranging from 0.8 to 1.2 $\rm M_{\odot}$, while in the case of He-accreting WDs the relevant mass range is 0.8~-~1.0~$\rm M_{\odot}$.
This result is important in the context of SNe Ia progenitors, as sub-Chandrasekhar CO~WDs accreting H/He rich material are predicted to leave more conspicuous imprints in the form of ionized, optically bright nebulae, which are more likely to be detected, than the ones  resulting from H-accreting, Chandrasekhar mass WDs. 
\\
{\bf iv)} Overall, the predicted line luminosities of the nebulae surrounding He accreting WDs are larger  than those surrounding H accreting WDs. This is because He accreting WDs have the tendency to be more luminous and hotter sources as the lower energy released per unit mass by He nuclear burning is counterbalanced by the higher accretion rates required for the He-accreting WD's steady accretion regime. 
\\
{\bf v)} Most of the nebular line luminosities show low sensitivity on the ambient medium density decreasing or increasing only moderately (depending on the H/He accreting WD's properties) as the density increases from 0.2 to 2 $\rm cm^{-3}$. A noticeable exception is the 
$[\ion{O}{III}]$~$\lambda$5007 line where an increase by one order of magnitude in the surrounding ISM density results in a strong enhancement by two to three times in its line luminosity. Thus, accreting WDs embedded in denser environments produce more bright $[\ion{O}{iii}]$ nebulae where in some cases -- e.g. for massive WDs accreting He-rich matter at high rates -- the $[\ion{O}{III}]$ 5007 \AA \ can be the  dominant emission line.  Finally, concerning the radial extent of the H and He ionized regions (as described by the parameters $R_{\rm H^{0},~0.5}$ and $R_{\rm He^{0},~0.5}$), the increase of the ISM density by one order of magnitude results in a drop of the relevant radii by about  a factor of four.

\begin{figure*}
\includegraphics[trim=0 0 0 0, clip=true,width=\textwidth,angle=0]{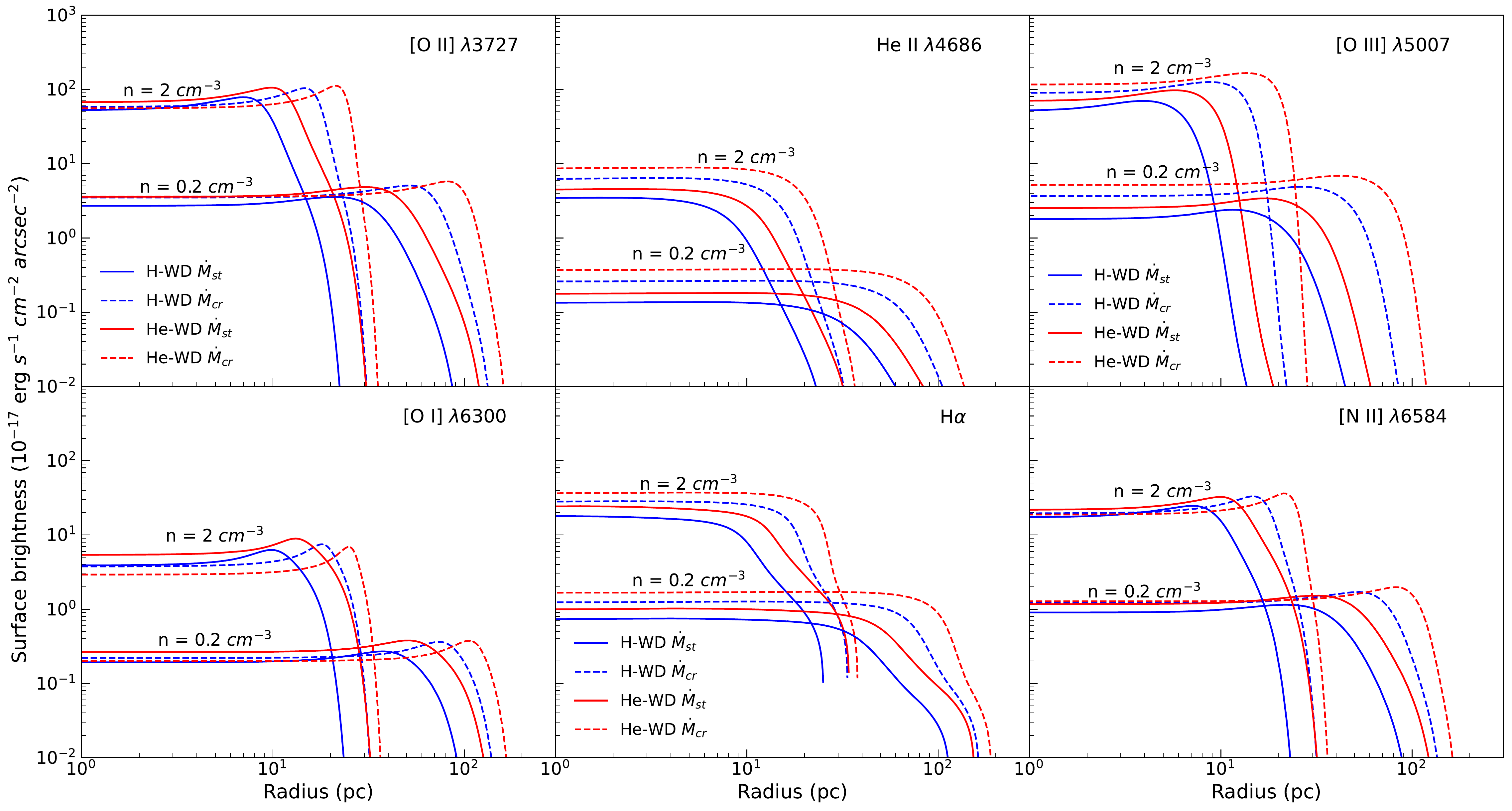} 
\caption {Comparison of surface brightness profiles for six prominent optical emission lines of nebulae ionized by a 1 $\rm M_{\odot}$ WD accreting H and He rich matter at the two corresponding limits of the accretion rate in the steady accretion regime ($\Dot{{M}_{\rm st}}$ and $\Dot{M}_{\rm cr}$). In each panel, the surface brightness profiles are shown for two constant ISM densities, 0.2 and 2 $\rm cm^{-3}$.} 
\label{fig:CSM_samplEedsd}
\end{figure*}

In order to clarify  how the observables of nuclear-burning WD nebulae depend on the  accretion rate, the chemical composition of the accreted material and the ISM density, we compare in Figure \ref{fig:CSM_samplEedsd}  the extracted  surface brightness profiles for six prominent optical emission lines of nebulae ionized by a 1~$\rm M_{\odot}$ WD, accreting H/He rich matter at the two corresponding limits of the steady accretion regime and for $n_{\rm ism}= 0.2$ and 2~$\rm cm^{-3}$. Even if, as mentioned above, the  line luminosities are not sensitive to the ambient medium density, the radial distribution of the nebular surface brightness changes substantially with the gas density.   As expected, decreasing the ISM density results in more extended, but fainter, ionized nebula formation.  In addition, as illustrated in our plots, WDs with higher accretion rates are expected to be surrounded by brighter and more extended ionized nebulae and thus, are more likely to be detected. The same applies for He-accreting WDs as they are hotter and brighter sources than their H-accreting counterparts,  and this has a direct impact on the size and the surface brightness of their surrounding ionized nebulae. Regarding the radial distribution of the surface brightness of each nebular line, the  $\ion{He}{II}~\lambda4686 $ and the $\rm H_{\alpha}$ lines reveal a rather flat curve, which after a given radius declines rapidly. By contrast, the surface brightness of the three oxygen forbidden lines ($[\ion{O}{ii}]~ 3727$~\AA, $[\ion{O}{iii}]~ 5007$~\AA \ and $[\ion{O}{i}]~ 6300$~ \AA)\ and the $[\ion{N}{II}]~6584$ \AA \ line initially increase moderately with increasing  distance from the source until a local peak can be seen after which it sharply declines. The radial position of this peak and the subsequent decrease of the surface brightness differs for each nebular line. Therefore, the nebular line ratios are expected to change substantially moving outwards from the ionizing source and hence, it is very important to consider the radial distance over which the line ratio has been obtained.  Intriguingly, the succession of local maxima in the nebula's surface brightness as a function of the radial distance from the source is in the same order  of the  flux peaks observed in the CAL 83 nebula, with the [\ion{N}{II}], [\ion{O}{I}] lines peaking further out than the [\ion{O}{iii}], $\rm H_{\alpha}$ and \ion{He}{II} lines. This result  supports the idea that the observed nebula in the vicinity of  CAL 83 is produced by the ionizing activity of the accreting WD.

Finally, we investigate the imprint of the nebulae around steadily-accreting WDs as compared to those resulting from other ionizing sources such as massive stars and accreting black holes by projecting our results in the phase-space of the so-called ``BPT''  diagrams \citep[][]{1981Baldwin}. These diagrams use  the [{O\,{\sc iii}}]/$\rm H_{\beta}$ vs. [{N\,{\sc ii}}]/$ \rm H_{\alpha}$, [{S\,{\sc ii}}]/$\rm H_{\alpha}$ and [{O\,{\sc i}}]/$\rm H_{\alpha}$ line ratios as an index that separates objects with different ionizing spectra. The line ratios of our models have been estimated by integrating the line luminosities over the entire nebula for solar gas abundances and for the two studied ISM densities (0.2 and 2 $\rm cm^{-3}$). In the same plots, we also include the theoretically/empirically derived relationships that indicate the regimes of the galactic Low-Ionization Emission-line Regions (LIERs), the Active Galactic Nuclei (AGN) and \ion{H}{II} region-like objects.

\begin{figure*}
\includegraphics[trim=0 0 0 0, clip=true,width=\textwidth,angle=0]{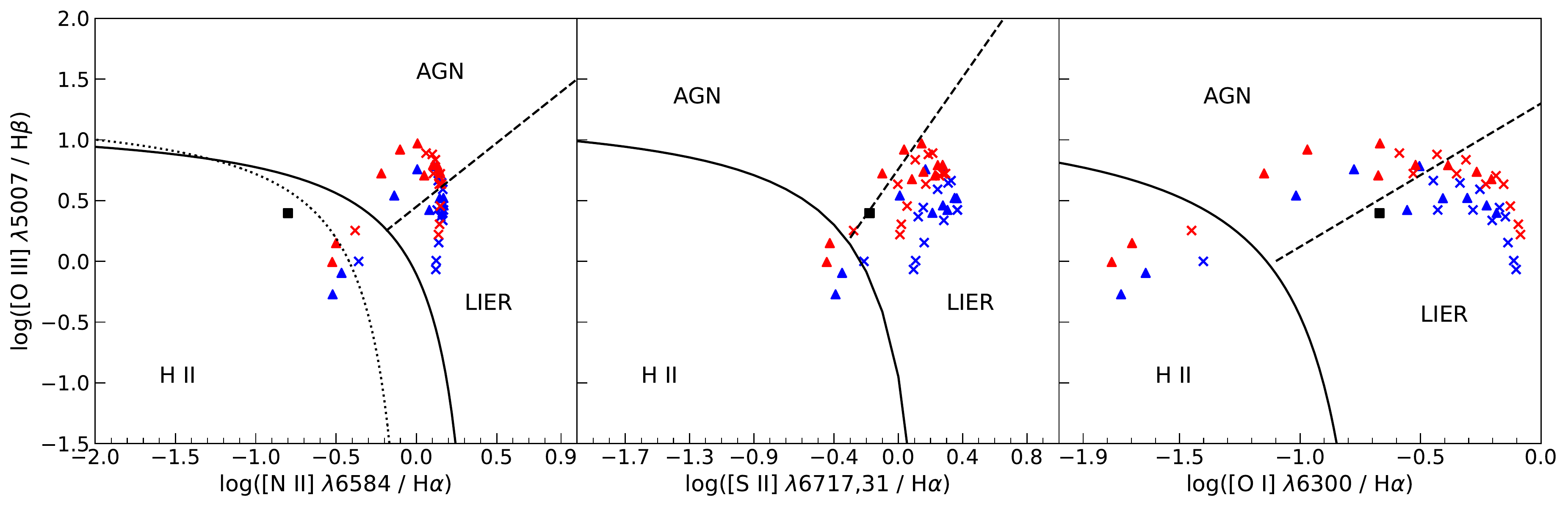} 
\caption { Diagnostic line-ratio diagrams 
for nebulae ionized by H-accreting (cross marks) and He-accreting (triangles) WDs compared with \ion{H}{II}, LIERs and AGN-like objects. The blue color corresponds to the ambient medium denisty of  $n_{\rm ism}= 0.2 \rm \ cm^{-3}$ and the red one to $ n_{\rm ism}= 2\rm  \ cm^{-3}$.  Left: [{O\,{\sc iii}}]/$\rm H_{\beta}$ vs. [{N\,{\sc ii}}]/$ \rm H_{\alpha}$ nebular line ratios. The solid curve is the theoretically modelled “extreme starburst line” \citet{2001ApJ...556..121K}, the dotted line shows the \citet{2003MNRAS.346.1055K} selection criteria, while the dashed line represent the AGN-LIER demarcation by \citet{2007MNRAS.382.1415S}. Middle and right: [{O\,{\sc iii}}]/$\rm H_{\beta}$ vs. [{S\,{\sc ii}}]/$\rm H_{\alpha}$ and [{O\,{\sc iii}}]/$\rm H_{\beta}$ vs. [{O\,{\sc i}}]/$\rm H_{\alpha}$, respectively. The solid and the dashed lines in the center and right diagrams represent the criteria of \citet{2006MNRAS.372..961K} used to separate \ion{H}{II}, AGNs, and LIERs. For comparison, we plot the estimated line ratios from a sub-region of the CAL 83 nebula (black square) as extracted by \citet{Gruyters2012}. }
\label{fig:regimesal}
\end{figure*}

As Figure \ref{fig:regimesal} illustrates, the nebular flux ratios around H-accreting WDs in the lower ionization lines [{N\,{\sc ii}}], [{S\,{\sc ii}}] and [{O\,{\sc i}}], relative to $\rm H_{\alpha}$, is much higher than the ionization by young massive stars ({H\,{\sc ii}} regions), while the [{O\,{\sc iii}}]/$\rm H_{\beta}$ line ratio is overall lower compared to the gas ionized by AGN-like objects.  One exception is the low-mass WD population with high accretion rates ($M_{\rm WD} \le 0.6 ~\rm M_{\odot}$); these   coincide  with the loci of  \ion{H}{II} regions. This result is due to the low photospheric temperatures ($T_{\rm eff}< 55000 \ \rm K)$ and luminosities ($\log(L/L_{\odot}) < 3.5 $) that characterise this subclass of  H-accreting WDs to possess ionizing properties comparable to those of massive OB stars. On the other hand,  for the high ISM density case ($n_{\rm ism} = 2~\rm cm^{-3}$), the rapidly-accreting WDs in the mass range of $M_{\rm WD} \sim 0.8 - 1.0 ~\rm M_{\odot}$, due to their high ionizing efficiency, produce enhanced [{O\,{\sc iii}}]/$\rm H_{\beta}$ line ratios and overlap with the AGN region. Nevertheless, apart from those exceptions, the vast majority of the nebulae around H-accreting WDs are well within the loci of LIERs.

The class of  He-accreting WDs, being  hotter and more luminous sources than their H-accreting counterparts, produce ionized nebulae characterised by higher [{O\,{\sc iii}}]/$\rm H_{\beta}$ ratios. Consequently, more members of this class are  found in the AGN regime. These are mainly the WDs with masses in the range of $M_{\rm WD} \sim 0.8 - 1.0 ~\rm M_{\odot}$ that steadily accrete He at the maximum possible rate ($\dot{M}_{\rm acc} \sim \dot{M}_{\rm cr}$). Similar to the previous case, the rapidly He-accreting, low mass WDs ($M_{\rm WD} \sim 0.6 - 0.8 ~\rm M_{\odot}$) are found within the \ion{H}{II} region. For the WDs that accrete He-rich material at low accretion rates, we find that there is not a great distinction with the H-accreting ones and are  also found in the general area of LIERs.

In the same diagrams we also plot  the relevant  observations of \citet{Gruyters2012}, who provide a spectroscopic study of a sub-region around CAL 83, the only known ionized nebula around a SSS. The observed flux ratio values [{O\,{\sc iii}}]/$\rm H_{\beta}$=2.50, [{S\,{\sc ii}}]/$\rm H_{\alpha}$=0.65 and [{O\,{\sc i}}]/$\rm H_{\alpha}$=0.21 of the nebula around CAL 83  coincide with the regions where our models are placed. This is not the case for the observed [{N\,{\sc ii}}]/$\rm H_{\alpha}$=0.15 ratio which is slightly lower  compared to our model's predicted range.  This discrepancy can be explained by the fact that CAL 83 is in the Large Magellanic Cloud, where the metallicities are probably smaller  by factor of 2-8, as well as by the fact that these line ratios arise from about one quarter of an inner region of the nebula, while our theoretical results concern the integrated luminosity of the entire nebula. As we have shown in this section, different regions of the nebula around accreting WDs possess different line ratios  \citep[see also][ for a similar argumentation]{Rappaport1994,remillard1995,Gruyters2012}. Further observations, covering the entire nebula of CAL 83 and a targeted modeling is required  to provide a better understanding on the properties of the ionizing source and its surrounding gas.

\section{Comparison with the observables of supersoft X-ray sources and type Ia SNRs}{\label{sec:3}}

Having determined the traces of accreting WDs on their surrounding gas, in this section we compare the results extracted by our modeling to the relevant constraints imposed by optical observations. In particular, we assess the compatibility of steadily accreting WDs with the constraints acquired by the lack of  (relic) ionized nebulae around a number of SSSs and SNRs~Ia \citep{Rappaport1994, kuuttila2019,Farias2020} as well as by the existence of Balmer dominated shocks found in many SNRs Ia \citep{woods2017, Woods2018}.

\subsection{Upper limits of \texorpdfstring{$[\ion{O}{iii}]$ 5007 \AA} ~ line luminosities enclosed within 6.8 pc of SSSs and SNRs Ia}

\begin{table*}


\caption{Properties of SNRs Ia considered in this study and constraints on their ambient medium. SNR parameters: Column 1: the name of each SNR; Column 2 and 3: the present radius in pc and age in years of each SNR; Column 4: logarithm of the luminosity's upper limit ($ \rm erg~s^{-1}$) enclosed within 6.8 pc of the central source; Column 5: upper limits on the $\ion{He}{ii} \ 4686$ \AA \ surface brightness for each SNR; Column 6: the minimum required neutral hydrogen fraction at the present radius of the shock of each SNR; Column 7: the pre-shock gas density range of each SNR as presented in the current literature.\\ References: (1) \citet{Rest2005}, (2) \citet{2014kosenko}, (3) \citet{Farias2020}, (4) \citet{kuuttila2019}, (5) \citet{warren2004}, (6) \citet{Ghahavian2002}, (7) \citet{2003ghavamian}, (8) \citet{Ghavamian2001}, (9) \citet{Ghavamian2007}, (10) \citet{Lewis2003}, (11) \citet{Yamaguchi2014}, (12) \citet{2006Badenes},  (13) \citet{Tian2011}, 
(14) \citet{2008konsenko}, (15) \citet{Williams2014}, (16) \citet{2007Raymond}, (17) \citet{2019Seitenzah}, (18) \citet{2003Rakowski}, (19) \citet{Acero2007}, (20) \citet{2013williams}, (21) \citet{Katsuda2010}.}

\label{tab:snrs}
\begin{tabular}{|c|c|c|c|c|c|c|c|}
\\ \hline
SNRs  & Radius  & Age     &  log($L_{\ion{O}{III},~6.8~\rm pc}$) & $\ion{He}{ii}~ 4686$~\AA \ Surface brightness & $f_{H^{0}}$ &  $n_{0}$  & References \\
&  (pc) & (years) & ($\rm erg~s^{-1}$) & $ (\times~10^{-19}~\rm erg~s^{-1}~cm^{-2}~arcsec^{-2}) $ & &  $(\rm cm^{-3})$\\ \hline
0519-69.0 & $4 \pm \ 0.3$ & $600 \pm \ 200$ & 34.08 & 5.3 & 0.4-0.5 &  1-2 & 1, 2, 3, 4, 9, 17 \\ 
0509-67.5 & 3.6 & $ 400 \pm \ 120 $ & 33.75 & 4.2 & >0.4 & 0.1-0.6 & 1, 2, 3, 4, 5, 9, 14, 17 \\
N103B & 3.6 & $860 \pm \ 400$ & unknown & 5.7 & unknown & 1-2.5 & 1, 4, 10, 15\\
DEM L71 & 6.8-9 & $\sim 4700 $  & unknown & 4.7 & 0.2-0.4 &  0.5-1.5 & 4, 7, 9, 18\\
SN 1006 & 10 &$\sim 1000 $ & unknown & unknown & >0.1 &  0.05-0.4 & 6, 11, 16, 19\\ 
Tycho & $ 3.3 \pm \ 0.3 $ &$ \sim 434 $ & unknown & unknown & >0.8 &  0.1-0.9 & 8, 12, 13, 20, 21 \\ \hline

\end{tabular}
\end{table*}

\begin{table}

\centering

\caption{Logarithmic upper limits of the $[\ion{O}{III}]~5007$~\AA \ luminosity ($\rm erg~ s^{-1}$) enclosed within 6.8 pc of the central source for the below SSSs, as extracted by the works of \citet{remillard1995} and \citet{Farias2020}. CAL 87 2 and CAL 87 3 refer to the observations of CAL 87 obtained at two different epochs.}
\label{tab:ssss}
\begin{tabular}{|c|c|}
\\ \hline
SSSs  &  log($L_{\ion{O}{III},~6.8~\rm pc}$) \\ \hline
RX J0550.0-7151 & 33.58 \\
RX J0513.9-6951 & 34.20 \\
CAL 87 2 & 34.01 \\
CAL 87 3 & 34.14 \\
Remillard+95 & 34.22 \\
\hline

\end{tabular}
\end{table}

\begin{figure}
\includegraphics[trim=7 0.3cm 9 8, clip=true,width=\columnwidth,angle=0]{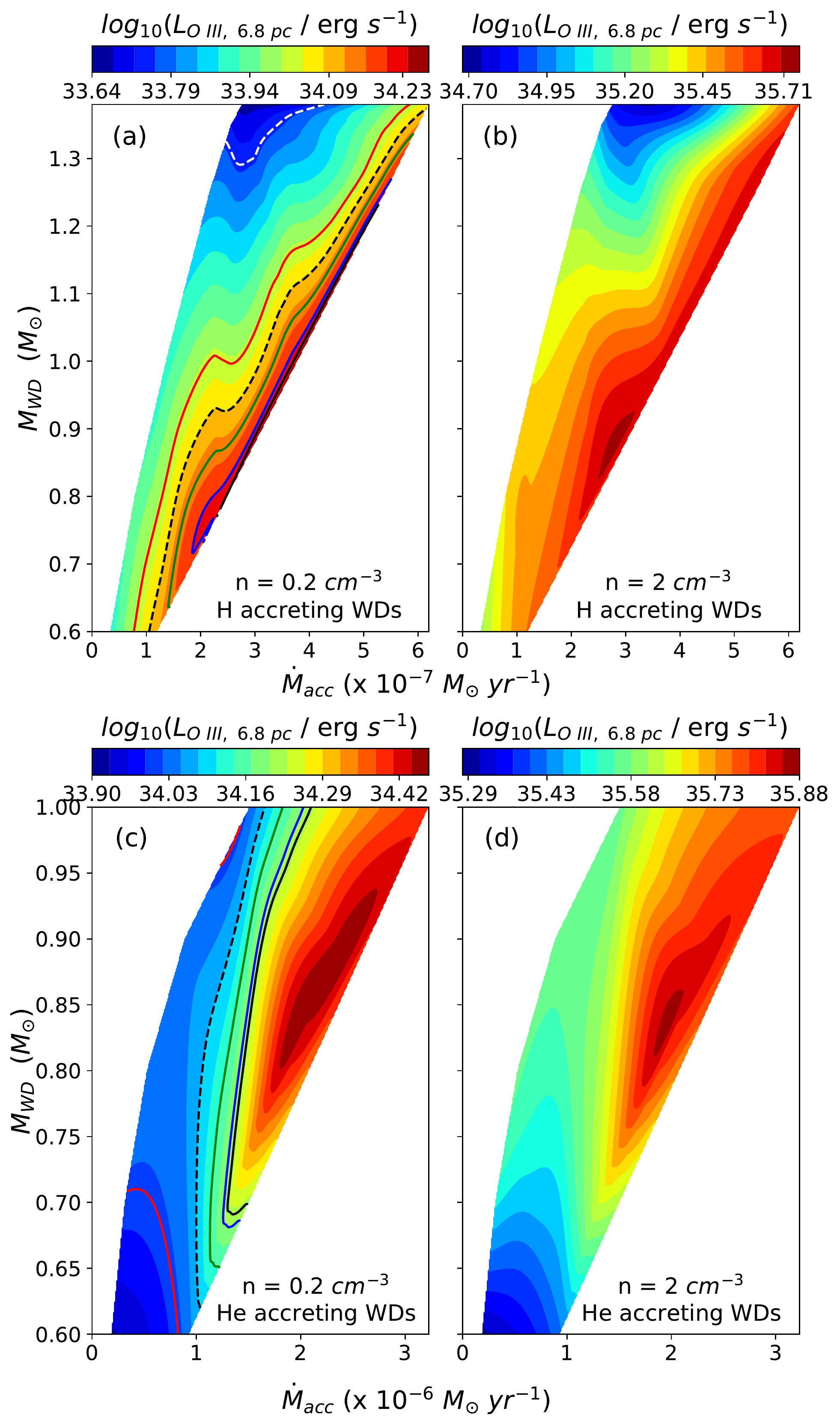} 
\caption {The $[\ion{O}{III}] 5007$ \AA \ luminosity enclosed within 6.8 pc radius as a function of the accretion rate and the WD mass in the steady accretion regime for: (a) H accreting WDs and constant ISM density $0.2~\rm cm^{-3}$, (b)  H accreting WDs and constant ISM density $2~\rm cm^{-3}$, (c)  He accreting WDs and constant ISM density $0.2~\rm cm^{-3}$ and (d) He accreting WDs and constant ISM density $2~\rm cm^{-3}$. The black and the white dashed lines represent the limits for SNRs 0519-69.0 and 0509-67.5, respectively. The black, blue, green and red solid lines 
represent the upper limits for the SSSs acquired by Remillard+95, RX~J0513.9-6951, CAL 87 3 and CAL 87 2, respectively. }
\label{fig:oiii}
\end{figure}

The ionization state of the ambient medium around SSSs and SNRs~Ia is well constrained by optical observations in the $[\ion{O}{III}] \ \lambda 5007 $ line, which is expected to be one of the dominant emission lines in the ionized nebulae around accreting WDs (see Sect. \ref{subsec:2.2}). The recombination timescale for doubly ionized oxygen  responsible for the forbidden line $[\ion{O}{III}] \ \lambda 5007 $ is $\sim 10^{4 }(n_{\rm ism}/1 \rm cm^{-3})^{-1} \rm yrs$ \citep{osterbrock2006}, so for any hot, luminous progenitor scenario, young SNRs would still be surrounded by these associated nebulae long after the explosion. 

\citet{remillard1995}  searched for  $[\ion{O}{III}]$ bright nebulae  around ten SSSs\footnote{The LMC targets in their study were: CAL 83, CAL 87, RXJ 0439.8-6809, RXJ 0513.9-6951, RXJ 0527.8-6954 and RX J0550.0-7151, while the SMC targets were: 1E 0035.4-7230, RXJ O048.4-7332, 1E 0056.8-7154 and RXJ~0058.6-7146.} in the Small and  Large Magellanic Cloud (SMC and LMC, respectively). They did not detect any optical emission  around these sources but only for one case, CAL~83. For the remaining SSSs, they set an upper limit of $10^{34.22}~\rm erg~s^{-1}$ on their $[\ion{O}{iii}]$ line luminosities enclosed within 6.8 parsec from the central sources \footnote{7.5 pc in their work, but they have assumed an LMC distance of 55 kpc.}.

Later, \citet{Farias2020} aiming to revisit some of the fields  studied by \citet{remillard1995} used advanced modern  instruments and they searched for nebulae around four of the studied LMC SSSs (CAL~83, CAL~87, RX~J0550.0-7151 and RX~J0513.9-6951). In addition, they extended their work to SNRs Ia, where they searched  for relic nebulae in the vicinity of three young LMC SNRs Ia:  N103B, SNR 0519-69.0 and SNR 0509-67.5. The authors confirmed that  no [$\ion{O}{iii}]$ ionized regions were detected around the studied SSSs and SNRs~Ia\footnote{The only exception was the SNR N103B, where they detected some emission, but they concluded that this emission does not correspond to a diffuse $[\ion{O}{iii}]$ ionized nebula around the source.}, except  for the SSS CAL 83, establishing more sensitive upper limits  on the $[\ion{O}{iii}]$ nebular line luminosities.  In Tables \ref{tab:snrs} (forth column) and \ref{tab:ssss}  the upper limits are summarized for the [$\ion{O}{iii}]$ ~5007~ \AA \ luminosity enclosed within 6.8 pc from the central source for the SNRs and the SSSs, as extracted by the works of  \citet{remillard1995} and \citet{Farias2020}.

In order to provide a direct comparison of our theoretical
predictions with the upper limits acquired from the observations, we follow the same methodological approach as described in Sect. \ref{subsec:2.1} and we calculate the nebular [$\ion{O}{iii}]$ line luminosity enclosed at 6.8 pc around H- and He-nuclear burning WDs. This time in our modeling we assume LMC-like chemical gas composition because the majority of our sample is in the LMC, with abundances one half the solar values \citep{choud2018}. The ambient medium number density in our models is taken to be $0.2~\rm cm^{-3}$ and $ 2~\rm cm^{-3}$,   values that can be considered as a  conservative lower and upper bound for the bulk ambient medium properties found around SSSs \citep[e.g.][]{Farias2020} and of young SNRs Ia \citep[e.g.][]{Yamaguchi2014,martinez2018}, and are  in line with the independent estimates of the pre-shocked densities of the remnants involved in this study (Table \ref{tab:snrs}, seventh column).

In Figure \ref{fig:oiii} we plot the upper limits of the [$\ion{O}{iii}]$ luminosity extracted by observations, together with the luminosity predicted by our models as a function of the WD mass and accretion rate,  for the two studied ISM densities. 
Our results clearly illustrate that the assumed ISM density around the  SSSs and SNRs substantially alters the constraints imposed by observations  in terms of the accreting WD properties. Specifically, for gas density equal to 2 $\rm cm^{-3}$, all of our models lie well above the derived upper limits for all the  SSSs and SNRs~Ia of our sample (see Fig. \ref{fig:oiii}b, d ). This means that if the ambient medium around these objects is  close to or higher than 2 $\rm cm^{-3}$ none of the  nuclear-burning  WD models are able to reproduce the relevant observables.

By contrast, when the density drops from 2 to 0.2 $\rm cm^{-3}$, we find that within our studied parameter space not all accreting WD models are excluded (Fig. \ref{fig:oiii}a,c). Specifically, we cannot rule out any sub-Chandrasekhar or Chandrasekhar mass WD accreting H/He rich matter with low or intermediate  accretion rates in the steady accretion regime for the SNR 0519-69.0. Stricter constraints are placed for the progenitors of SNR 0509-67.5, where only a Chandrasekhar mass WD accreting H with low rates could be a potential progenitor, while all He-nuclear burning WDs are ruled out.  In the case of SSSs, the upper limits acquired from \citet{remillard1995} and \citet{Farias2020} are consistent with low- and intermediate-accretion rates for both H and He  accretors, and hence, the non-detection of [$\ion{O}{iii}]$ ionized regions around  these sources can potentially be explained within the framework of nuclear-burning WD models embedded in an ambient medium with density close to or lower than $0.2~\rm cm^{-3}$. The only exception is RX~J0550.0-7151, where none of the models can explain the lack of [$\ion{O}{iii}]$ line, which that indicates that the source should be surrounded by an ISM with density substantially  smaller than 0.2~$\rm cm^{-3}$.

 \begin{figure*}
\includegraphics[trim=60 10 250 20, clip=true,width=\textwidth,angle=0]{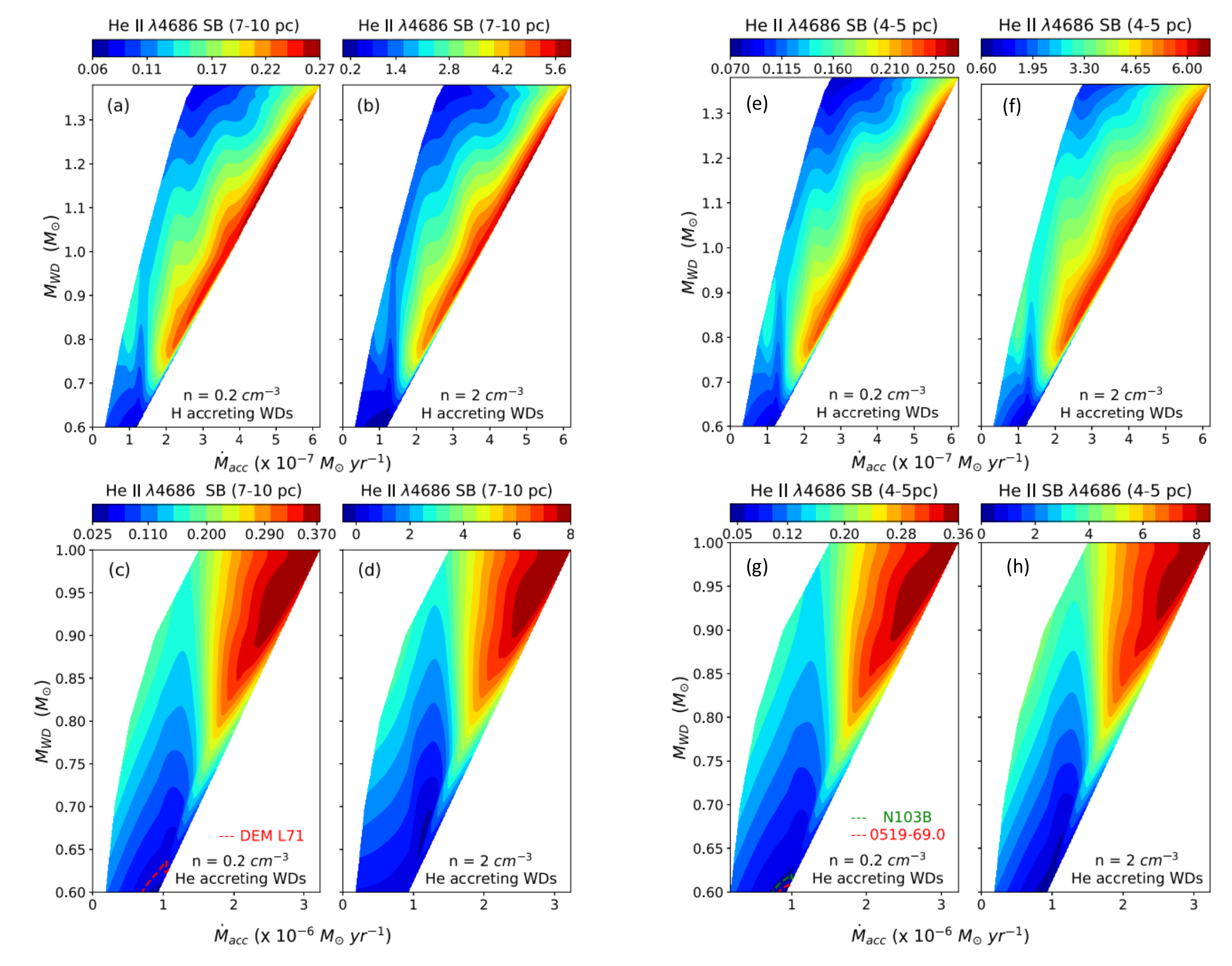} 
\caption { The averaged  $\ion{He}{II}~4686$ \AA \ surface brightness in units $\times~10^{-17}~\rm~erg~s^{-1}~cm^{-2}~arcsec^{-2}$ at 7-10~pc (a-d) and 4-5~pc (e-h) from the ionizing source, as a function of the accretion rate and the WD mass. The dashed line at (c) represents the upper limit of DEM L71  and in (g) those of 0519-69.0 and N103B as extracted by \citet{kuuttila2019}. In the rest plots the upper limits extracted by observations are bellow the range extracted by our models. }
\label{fig:HeII}
\end{figure*}

\subsection{Upper limits on the  \texorpdfstring{$\ion{He}{ii}$  4686 \AA} ~  surface brightness for SNRs Ia}
  
Additional constraints on the ionizing flux of the SNe Ia progenitor systems have been placed by  the IFU observations carried out by \citet{kuuttila2019}, who searched for $\ion{He}{ii}$ 4686 \AA \ nebular emission around the three  SNRs studied above (0509-67.5, 0519-69.0 and N103B) and the LMC SNR DEM L71. The observations  resulted in a non-detection and thus, strict limits were extrapolated on the $\ion{He}{ii}$ 4686 \AA \ nebular fluxes at the vicinity of the studied remnants (see fifth column of Table \ref{tab:snrs}).

Here, we apply the same method as \citet{kuuttila2019} to compare our results extracted by {\sc cloudy} with their acquired upper limits. For that reason, we estimate the average surface brightness of the $\ion{He}{ii}$ 4686 \AA \  emission line in  an annulus 4--5~pc around  the SNRs 0509-67.5, N103B and 0519-69.0, and  7-10~pc around the SNR DEM L71, which correspond to the area ahead of the forward shock in each remnant.  In Figure~\ref{fig:HeII}, we plot the upper limits of the $\ion{He}{ii}$~4686~\AA \ surface brightness for each object together with the average surface brightness in the range of 7-10 pc (Fig. \ref{fig:HeII}a-d) and 4-5~pc (Fig. \ref{fig:HeII}e-h),  predicted for the H/He nuclear burning WD models as functions of WD mass and accretion rate and for the two constant ISM densities of 0.2 and 2 $\rm cm^{-3}$.

As is evident from  Figure~\ref{fig:HeII}, almost all the H/He accreting WD models lie well above the derived upper limits for the four SNRs in both density cases of 0.2  and 2 $\rm cm^{-3}$. There is only a small window where the upper limits for SNRs DEM L71, N103B and 0519-69.0 are slightly higher than our results  in the low density case and for WD masses $\sim 0.6~\rm {M_{\odot}}$ accreting He-rich matter with high accretion rates (see Fig~\ref{fig:HeII} c and g).  Nevertheless, WDs with this mass  cannot produce SNe Ia and thus, are not of interest  here.

In conclusion, even if the limits  imposed by the lack of nebular [\ion{O}{iii}] allow some accreting  WDs systems to be placed as potential progenitors of the studied SNRs Ia, the  non-detection of the \ion{He}{ii} nebula emission  rules out essentially any H/He steadily nuclear-burning WD system in which the accretion process was occurring for about $7 \times 10^{4}$ years prior the explosion \citep[i.e. the recombination timescale for helium;][]{Pequignot}. Alternatively, these remnants  could be surrounded by ISM densities well below 0.2~$\rm cm^{-3}$, but such a result  is in conflict with their current dynamics and emission properties \citep{Yamaguchi2014,martinez2018}.

\subsection{Upper limits from the existence of Balmer dominated shocks in SNRs Ia }

\begin{figure*}
\includegraphics[trim=0cm 0.3cm 0cm 0cm, clip=true,width=\textwidth,angle=0]{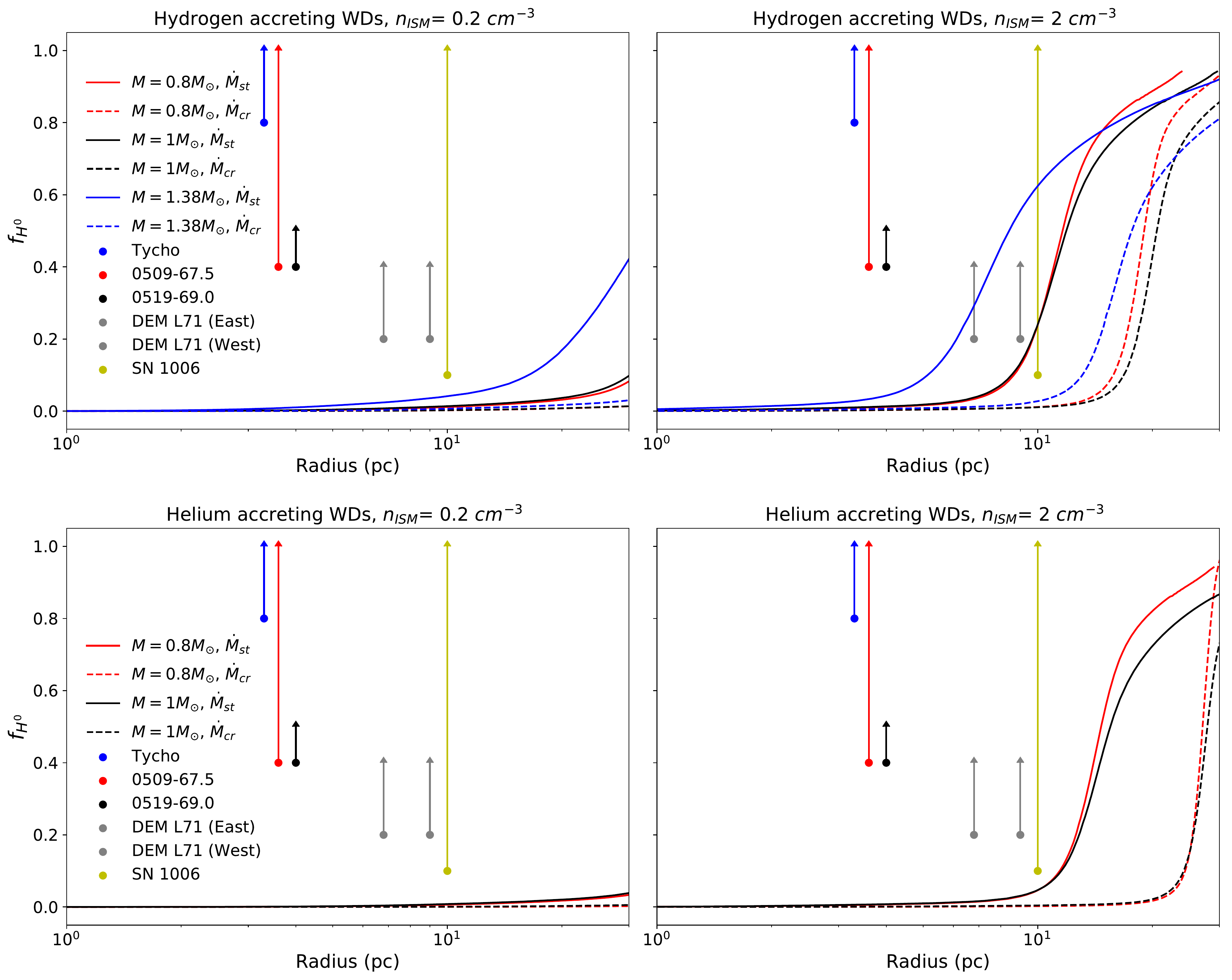} 
\caption {Hydrogen neutral fraction as a function of radial distance from CO WDs accreting H and He rich matter at the two stability boundaries in the steady accretion regime for various WDs' masses. The minimum hydrogen neutral fractions at the present radius of the shock of each remnant are also plotted. }
\label{fig:balmer}
\end{figure*}

Another restriction on the progenitors of SNe Ia comes from the existence of Balmer dominated shocks that are frequently found in SNRs that result -- or are suspected to result -- from SNe Ia. Their optical spectra show regions that exhibit Balmer line emission with a narrow core and a broad base along the forward shock.  Balmer dominated shocks arise when the forward shock moves into a partially neutral ISM \citep{chevalier1980}. The narrow component is understood to result from collisionally excited neutrals, while the broad component is due to charge exchange \citep[][and references therein]{Raymond2001,Heng2010}. As a consequence, the existence of at  least some neutral hydrogen in the vicinity of SNRs Ia limits the ionization history of their progenitors  to about $\tau_{\rm rec} \sim 10^{5}$ years prior to the explosion \citep{woods2017, Woods2018}.

In Table \ref{tab:snrs} (sixth column), we present the minimum required neutral hydrogen fractions,  which are  consistent with observations and numerical models of the Balmer-dominated shocks \citep{Ghahavian2002,2003ghavamian} associated with two SNRs Ia in our Galaxy (Tycho and SN 1006) and three LMC SNRs Ia (0519-69.0, 0509-67.5 and DEM~L71). 

Aiming to assess the limitations imposed by the Balmer shocks on SNe Ia progenitors, in Figure \ref{fig:balmer} we present the minimum hydrogen neutral fractions estimated for each SNR Ia of our sample  versus its present radius, together with the hydrogen neutral fraction as a function of radial distance from H/He nuclear burning WDs at various WDs masses that accrete matter at the two stability limits in the steady accretion regime as extracted by our models.

As Figure \ref{fig:balmer} clearly displays, adopting a surrounding ISM density of 0.2 $ \rm cm^{-3} $, any steadily nuclear burning activity on their WD progenitors is  ruled out for $ \sim 10^{5} $ years prior the explosion. This result is independent on the assumed WD mass and accretion rate. In the highest density case ($n_{\rm ism}= 2~\rm cm^{-3}$), He nuclear-burning WDs continue to remain inconsistent with the limits acquired from observations, while in the case of H-rich accreted material, WDs accreting matter with low accretion rates cannot be excluded as potential progenitors of SNRs DEM L71 and SN 1006. However, SN~1006 is known to be expanding into much lower density medium  (see seventh column of Table \ref{tab:snrs}) and particularly, its pre-shock gas density has been estimated to be $n_{\rm ism}= 0.25 - 0.4 ~\rm cm^{-3}$ at the northwestern, Balmer-emitting quadrant of the remnant \citep{2007Raymond}.

Summing up, the existence of Balmer dominated shocks rules out almost any possible stable nuclear burning WD model (both sub- or Chandrasekhar-mass) that was accreting matter at least~ $\sim 10^{5}$~years prior the SNe Ia explosion. Alternatively, a surrounding ISM  with densities equal to or higher than 2 $\rm cm^{-3}$ is required.  Such a result leads to a fatal controversy to the restrictions imposed by the lack of \ion{He}{II} and [\ion{O}{iii}] bright nebulae around these SNRs as the latter demand  ISM densities close to or lower than  0.2 $\rm cm^{-3}$. Conclusively, any steadily nuclear-burning WD, embedded in a constant ambient medium, seems incapable to satisfy simultaneously the overall observables collected for the studied sample of SNRs and thus, is excluded as a potential  SN Ia progenitor for these sources. Such a conclusion stands for any given WD mass and accretion rate and thus, strong limitations are imposed for both the Chandrasekhar and sub-Chandrasekhar mass, single degenerate models of SNe Ia.

\section{Discussion}
\label{sec:4}

\subsection{The imprints of accreting WDs on their ambient medium}

In this work, aiming to put further insights into the interaction of accreting WDs with their ambient medium, we modeled the emission line spectra of the nebulae around steadily H and He nuclear-burning WDs by pairing known WD accretion models with radiation transfer simulations.  Having included in our modeling a large range of CO WD masses (0.6~-~1.38~$\rm M_{\odot}$ for H-accreting WDs and 0.6~-~1.02 $\rm M_{\odot}$ for He accreting WDs), all possible accretion rates within the steady accretion regime and plausible ISM densities, we studied how the ionization state of the gas and the optical nebular emission lines  vary with the aforementioned parameters.

Our results show that the  level of ionization, the  radial extent and the spectral properties of the formed ionized nebulae  vary strongly with the properties of WD accretion. In particular, we found that the   nebulae are expected to be brighter (by about one order of magnitude) and more extended  (by a factor of two) when the mass accretion rate increases from its minimum to its maximum value for a given WD mass. The nebular emission lines are also heavily dependant on the WD mass, where the most extended and brighter nebulae are expected around intermediate WD masses  ($M_{\rm WD} \sim 0.8 - 1.2~\rm M_{\odot}$). Regarding the chemical composition of the accreted material, we found that He-accreting WDs ionize their ambient medium  to larger distances and their nebular emission lines are brighter compared to their H-accreting  counterparts. 

These results advocate that a vast diversity of properties is expected to be met in the medium around steadily-accreting WDs, as each nuclear-burning WD leaves its unique and distinctive imprint on the surrounding gas. This is essential  in the hunt for accreting WD systems through the detection of their ionized nebulae,  given that the optical properties of the ionized gas reveal encoded information about the the central source. Our work offers  diagnostic tools for this information to be deciphered.

Given that accreting WDs  are expected to be numerous in every galactic type \citep{VandenHeuvel1992, Rappaport1994a}, it is very unlikely CAL 83 is the only detected case of a nebula shaped by a WD's accretion activity.  Many ``orphan'' (without identified ionizing sources), extended ionized nebulae, such as HBW~ 673 and BCLMP~ 651 identified in M33 \citep{2011kehrig, 2016Woods} reveal similar properties  to those obtained in this work and, thus, they represent ideal cases to be investigated  as possible accreting WD systems. Overall, in agreement with previous studies, we showed that the ideal optical line to search for  nebulae shaped by accreting WDs is the forbidden line [\ion{O}{III}]~ 5007~ \AA; we predict this line to be the brightest optical line of the nebulae around most accreting WDs.  Nevertheless, our models included the exploration of the forbidden [\ion{O}{II}] 3727~\AA \ line that had not been studied previously. We found that in low density media ($n_{\rm ism}~ \le ~0.2~\rm cm^{-3}$) the [\ion{O}{III}] 5007~ \AA \ line is very reduced and the most prominent line becomes the [\ion{O}{II}] 3727 \AA.  Thus, this line could potentially be useful in observational searches for accreting WD nebulae, especially if these sources are embedded in low ambient medium densities.  Moreover, bright and dominant optical signatures of the shaped nebulae around accreting WDs are expected also to be the  $\rm H_{\alpha}$, [\ion{N}{II}] 6584 \AA \ and [\ion{S}{II}]  6717 \AA \ emission lines as they are the other brightest  lines among the  emission lines studied in all of our models. Finally, regarding the likelihood of accreting WD detection through their ionized nebulae, our modeling has shown that intermediate mass WDs ($M_{\rm WD} \sim 0.8 - 1.2~\rm M_{\odot}$) that accrete steadily  at the maximum possible range ($\dot{M}_{\rm acc} \sim \dot{M}_{\rm cr}$) have  a higher probability to be identified because they produce the most extended and bright nebulae. The same applies for the He accreting WDs (compared to their H accreting counterparts) since they have the tendency to be hotter and more luminous ionizing sources.

To further untangle the expected imprints of steadily-accreting WDs compared to other ionizing sources, we explored how the extracted nebular line emission ratios fit to various diagnostic diagrams used to separate and characterize emission line objects.  Our findings agree with those extracted by  \citet{Rappaport1994}, showing that the nebulae around accreting WDs are clearly distinctive from those of typical \ion{H}{II} regions. In particular, all nebulae produced by our models possess [\ion{N}{II}] $\lambda 6584$, [\ion{S}{II}] $\lambda 6717,31$ and [\ion{O}{I}] $\lambda 6300$  over $\rm H_{\alpha}$ ratios that are much higher than the ionization by young massive stars. However, there is some contamination between the properties of the two objects where for the case of highly-accreting, low-mass WDs ($M_{\rm WD} \le 0.6~\rm M_{\odot}$) the formed  nebulae are expected to share similar properties to those of \ion{H}{II}-like objects.  \citet{Rappaport1994} focused on hot, luminous accreting WDs -- expected to be the members of the SSS class --  and argued that the [\ion{O}{III}] $\lambda 5007$/$\rm H_{\beta}$ ratio can be used as a distinctive diagnostic tool between the nebulae around massive accreting WDs and \ion{H}{II} regions, as the line ratio values in their supersoft models were much higher relative to those in \ion{H}{II} regions. Nevertheless, considering the whole range of accreting WDs, this argument is no longer valid as we found that the ranges of the [\ion{O}{III}] $\lambda 5007$/$\rm H_{\beta}$ ratio for the \ion{H}{II} regions and steadily nuclear-burning WD nebulae overlap significantly. Thus, the lower ionization lines [\ion{N}{II}] $\lambda 6584$, [\ion{S}{II}] $\lambda 6717,31$ and [\ion{O}{I}]~ $\lambda 6300$ compared to $\rm H_{\alpha}$ seems a safer diagnostic tool among the two classes of ionized nebulae. Another distinctive consequence of our broader and more realistic modeling is the extraction of a much wider range for nebula line ratios than previous works inferred. This is important in the context of ionized nebulae around accreting WDs, because the previous theoretical SSS models predict for most of the ions much higher line ratios (compared with $\rm H_{\beta}$) than \citet{Gruyters2012} observed and they did not have a firm explanation for these systematic discrepancies.  The response of the WD to mass accretion seems able to account for such variances. Some examples include the [\ion{O}{III}]/$\rm H_{\beta}$ and [\ion{N}{II}]/$\rm H_{\beta}$ ratios, which are much lower in our models and in a better agreement with  CAL~83 observations.

Finally, an additional interesting finding  derived from the BPT diagrams of our models is that the vast majority of the nebulae around H and He nuclear burning WDs coincide to the loci of the galactic low-ionization emission-line regions (LIERs). The origin of LIERs had been attributed to the ionization produced by the activity of a low-luminosity AGN \citep{1983ApJ...264..105F}. However, the  discovery of spatially extended low-ionization emission-line regions \citep{1996A&AS..120..463M, 2016A&A...588A..68G}, implies ionization  by an extended component that follows a stellar-like radial distribution \citep{2013A&A...558A..43S}. Until now, it has been suggested that the only viable extended source of ionizing photons for passively evolving galaxies are the hot post-asymptotic giant branch (AGB) stars \citep{2013A&A...558A..43S,2019AJ....158....2B}. Our models indicate that steadily-accreting WDs -- especially those that accrete H-rich material -- could be an alternative/additional source of LIERs as they seem to produce nebulae that share the same spectroscopic properties.  Similar to post-AGB stars, accreting WDs are expected to be hosted in copious amounts in galaxies of all morphological types,  which can potentially explain why LIERs are mostly found in galaxies with old populations and display little or no star formation activity. On the other hand, given that the accreting WD's  lifespan can be orders of magnitude larger than the short-lived post-AGB phase makes the former more persistent and thus, more efficient ionizing sources.  The same applies for the Diffuse Ionised Gas \citep[DIG,][]{Reynolds1984,Haffner2009} and especially for its HOLMES (hot low-mass evolved stars) component whose emission flux ratios are nearly always consistent with the LIERs loci in the BPT diagrams \citep[e.g.][]{Kumari2019}. Steadily-accreting WDs could potentially contribute to the formation of these nebulae together with the post-AGB stars that up to date have been proposed to be their ionizing source \citep[e.g.][]{2019AJ....158....2B}. However, even if this  explanation is tempting, in order to draw a firm conclusion on the contribution of steadily-accreting WDs on the  LIERs/DIG formation, detailed radiation transfer simulations coupled with binary population synthesis techniques are required.

\subsection{Comparing the results of our modeling to the relevant observations}

In the second part of this paper, we compared the results extracted by our radiation transfer simulations to the relevant observables we receive from the vicinity of accreting WD systems and related objects. The aim of this comparison was to (re)assess -- with broader and more detailed modeling -- whether the theoretical predictions on the ionizing flux from steadily-accreting WDs are aligned with the lack of any ionized nebulae around the most massive members of their class, i.e. the SSSs, and subsequently to investigate if steadily-nuclear burning WDs could be the progenitors of a number of well-studied SNRs Ia. 

Regarding the SSSs, our modeling has shown that the non-detection of [\ion{O}{III}] bright ionized regions around the nine studied SSSs in the LMC/SMC establishes an upper limit for their ambient medium densities equal or close to 0.2~$\rm cm^{-3}$, while for the case of the SSS  RX J0550.0-7151 a lower density ambient medium is required. These constraints on the ISM density around the studied SSSs are the strictest compared to the relevant ones of the current literature \citep{woods2017,Farias2020} and are applied for any possible combination of WD mass and accretion rate within the stable accretion regime.   

The demand of an ambient medium with density around the studied SSSs well below than 0.2 ~$\rm cm^{-3}$  is quite contradicting with the general expectations, given that the typical range of LMC midplane densities is $n_0~\approx~ 0.1~ -~ 4~  \rm cm^{-3}$ while the inferred ISM densities around these sources have been estimated to be $n_{\rm sss}= 0.22 - 2.26~ \rm cm^{-3}$, based on the total \ion{H}{I} column density in the direction of each source and assuming an isothermal disk distribution \citep[][]{2016Woods}. However, hot and very low density ISM densities are met in the LMC/SMC and are linked to the hot supernova-heated component \citep{Cox2005}. 
Thus, there is still the possibility of a ``natural coincidenc'' under which all studied SSSs,  except CAL 83, are embedded in the most tenuous regions of their host galaxy.  

This scenario is not applicable for the case of the studied SNRs~Ia. The observational pieces of evidence involved in our study,  i.e. the lack of nebular [$\ion{O}{III}]$ and $\ion{He}{II}$ emission around the SNRs Ia and the existence of Balmer lines in their shock waves, work complementary to each other in excluding stably-accreting WDs as possible progenitor systems. In particular, the absence of detected [$\ion{O}{III}]$ and $\ion{He}{II}$
nebulae in the vicinity of SNRs Ia excludes all stable nuclear burning WDs for ISM densities of $n_{\rm ism} > 0.2~\rm cm^{-3}$ while the existence of Balmer dominated shocks are inconsistent for ISM densities  of $n_{\rm ism} < 2~\rm cm^{-3}$.  Thus, within the steady accretion regime and for both H- and He-accreting WDs, there is not any possible combination of WD mass, accretion rate and ISM density capable to reproduce simultaneously the two observational facts that the SNRs Ia impose. This result essentially rules out any WD that was steadily-burning the accreted matter $\sim 10^{5}$ yrs before its final explosion as a possible progenitor of the studied remnants, and therefore, challenges both the Chandrasekhar and sub-Chandrasekhar mass  scenarios arising from single degenerate progenitors. Inevitably, the double degenerate scenario is placed as the most plausible origin of the studied remnants. 

Nevertheless, such a bold statement is  in possible tension with a number of findings that indicate that some of the studied SNRs might have originated from a single degenerate SN Ia \citep[see][for an extensive discussion and references therein]{Maoz2014, 2018Livio} and thus, before drawing a final conclusion,  some possible caveats need to be addressed in order to achieve a holistic picture that emerges  by combining the observations and the  different, relevant theoretical predictions.
Some of them, already discussed in the literature, refer to a ``time window'' between the WD accretion phase and the final SN Ia explosion. If this time interval is larger than the recombination timescale of the surrounding nebula then at the moment of the explosion the surrounding gas will have recombined returning into its (partially) neutral state. This time delay can be achieved through the so called ``simmering phase''  \citep{2016martinez,2017piersanti}, the ``spin up/spin down'' WD model \citep{diistefano2011}, as well as,  in the sub-Chandrasekhar edge-lit detonation models, where the WD spends a substantial time in the helium detonation regime (see Fig. \ref{fig:regimes}) before its fatal explosion \citep{Ruiter2011}.  All the above scenarios, plausible or not, could offer the desired time delay as long as mass accretion is ceased during this period.

Another possible parameter that we could question is whether the accreting WD progenitors were indeed embedded in a homogeneous ambient medium.  Given that SNe Ia result from  intermediate mass stars, no substantial modifications are expected to occur from their progenitor systems. Indeed, the overall morphological \citep{Lopez2009} and X-ray properties \citep{Badenes2007, Yamaguchi2014} of SNRs Ia indicate that these objects are evolving in a rather homogeneous ambient medium.  However, detailed studies of individual SNRs Ia have shown that their explosion centers were surrounding by moderate circumstellar structures and currently the remnants are interacting with them or they had an interaction history with them in the recent past. Such cases are for instance the young remnants of Kepler’s SNR \citep{chiotellis2012, Burkey2013, Toledo-Roy2014}, Tycho’s SNR \citep{Dwarkadas1998, Katsuda2010, Chiotellis2013}, G1.9+0.3 \citep{Borkowski2013, Borkowski2017, 2021chiotellis}, RCW86 \citep{Williams2011, Broersen2014}, N103B \citep{Li2017, Yamaguchi2021}  and the more evolved ones of DEM~L238, and DEM L249 \citep{Borkowski2006} etc.  Recently, the detection of numerous high density knots around the LMC remnants  0519-69.0, N103B, DEM L71, and 0548-70.4 are also associated with a CSM ejected by the SN progenitor system \citep{LI2021}.  Such circumstellar structures can be formed by the mass outflows of the progenitor system as it is evolving towards a SN Ia. The formation of the WD itself requires the ejection of its hydrogen rich envelope during the planetary nebula phase or by a common envelope episode. The ejected envelope itself is capable to shape extended circumstellar structures around the binary system with  lifetime up to several Myr \citep{2020chiotellis}.  In addition, the ionizing flux produced by accreting WDs leads to an overpressure of the heated ambient gas that results  in the expansion of surrounding nebula and the formation of non-homogeneous bubbles around the binary system \citep{Rappaport1994}.  A modified, non-homogeneous ambient medium around the accreting WD will substantially alter the ionization and emission properties of the resulted nebula and thus, it can serve as an additional possible solution for the lack of an ionized region around the studied SNRs Ia. Note that this scenario is also applicable for the non-detection of ionized nebulae around the studied SSSs for which the ``time window'' scenario discussed above  is not valid because for those objects, accretion is an ongoing process.

Nevertheless, detailed radiation transfer modeling is required to assess if indeed such circumstellar structures can offer a plausible explanation of the current observables we receive from SSSs and SNRs and if so, to identify the properties and the origin of these structures (Souropanis et al. in prep).

\section{Summary}
\label{sec:5}

In this paper we studied the observational imprints that steadily-accreting WDs leave to their ambient medium due to their ionizing activity. By coupling known H- and He-accreting WD models with radiation transfer simulations we extracted the sizes, the optical line luminosities and the surface brightness profiles of the surrounding ionizing gas as a function of the WD mass, the accretion rate, the ISM density and the chemical composition of the accreted material.

Our results have shown that all accreting WDs  are efficient ionizing sources capable to form extended, optical bright nebulae in their vicinity. The most efficient sources were found to be the intermediate mass accreting WDs ($M_{\rm WD} \approx 0.8 -1.2~\rm M_{\odot}$) that accrete mass in the maximum possible rate of the steady accretion regime.  These sources form the largest and brightest nebulae around them and thus, are more likely to be detected. The same applies for He-accreting WDs as, compared to their H-accreting counterparts, they appear to be hotter and more luminous sources and thus, more efficient ionizing sources. Overall, the ionization and emission properties of the shaped nebulae are strongly dependent on the accretion properties of the central WD. That means that each accreting WD for a given mass and accretion rate shapes an ionized nebula around it characterized by unique and distinctive properties.  Thus, through our modeling we provide direct links through which the emission properties of the surrounding nebula can be encoded revealing the nature and the properties of the central source.  

The formed ionized nebulae represent efficient tools for detecting steadily-accreting WDs as in most cases the source itself is utterly obscured. Among the studied recombination and forbidden lines, we found that the most luminous ones are  [\ion{O}{III}] 5007 \AA \ and [\ion{O}{II}] 3727 \AA;  the latter being the dominant one in low density environments. Thus, these two oxygen lines  are ideal traces for the detection of ionized nebulae around accreting WDs.  Other lines in which the ionized gas around steadily-accreting WDs is expected to be bright are those of $\rm H_{\alpha}$, [\ion{N}{II}] 6584 \AA \ and [\ion{S}{II}]  6717 \AA.

We  displayed our results on the so-called ``BPT'' diagrams and we found that the nebulae formed by accreting WDs reveal  properties that are distinct from  \ion{H}{II}-like regions, possessing higher [\ion{N}{II}] $\lambda 6584$, [\ion{S}{II}] $\lambda 6717,31$ and [\ion{O}{I}] $\lambda 6300$  over $\rm H_{\alpha}$ ratios. However, some contamination in the line ratios exists between the two objects for the case of highly-accreting low-mass WDs ($M_{\rm WD} \le 0.6~\rm M_{\odot}$). Intriguingly, the majority of the nebulae formed by steadily nuclear-burning WDs  share the same optical line ratio properties as those observed in galaxies that exhibit low-ionization emission-line regions (LIERs). We suggest that accreting WDs, being numerous in any galactic type, can potentially be one of the main contributors responsible for the formation of LIERs. 

Finally, we compared the results extracted by our theoretical modeling to the relevant observations conducted in the vicinity of a number of SSSs and SNRs Ia. We found that the non-detection of any ionized optical  nebula around all studied SSSs,  with the exception of CAL 83, establishes an upper limit on their ambient medium density equal to 0.2 $\rm cm^{-3}$. Regarding the studied SNRs, we found that the lack of nebular [$\ion{O}{III}]$ and $\ion{He}{II}$  emission in the vicinity of their forward shocks in combination to the existence of Balmer dominated lines mutually exclude any accreting WD progenitor that was steadily burning the accreted material $t \sim 10^5$~yrs before the SN Ia explosion. This result concerns any single degenerate accretion model including the  canonical Chandrasekhar mass models and the sub-Chandrasekhar (WD +  helium-rich donor star channels).  We discuss possible alternatives that can potentially eliminate  any possible discrepancies between the  predictions of the single degenerate models and the relevant observables. We claim that a plausible solution would be the consideration of a non-homogeneous ambient medium around the SN Ia explosion center, modified either by the mass outflows of the progenitor system or/and by the ionizing radiation produced by the accreting WD.

\section*{Acknowledgements}

The authors would like to thank the referee for  their thorough
comments that improved the manuscript. This research is co-financed by Greece and the European Union (European Social Fund-ESF) through the Operational Programme “Human Resources Development, Education and Lifelong Learning 2014-2020” in the context of the project “On the interaction of Type Ia Supernovae with Planetary Nebulae” (MIS 5049922). A.C. acknowledge the support of this work by the project ``PROTEAS II'' (MIS 5002515), which is implemented under the Action ``Reinforcement of the Research and Innovation Infrastructure'', funded by the Operational Programme ``Competitiveness, Entrepreneur- ship and Innovation'' (NSRF 2014–2020) and co-financed by Greece and the European Union (European Regional
Development Fund).  M.C. acknowledges support by NSF (1910687), NASA (19-ATP19-0188), and STScI (HST-AR-14556.001-A). L.P. acknowledges partial financial support from the INAF-mainstream
project “Type Ia Supernovae and their Parent Galaxies:
Expected Results from LSST. 
A.J.R. is supported by an Australian Research Council Future Fellowship through award number FT170100243.  G.F. acknowledges support by NSF (1816537, 1910687), NASA (ATP 17-ATP17-0141, 19-ATP19-0188), and STScI (HST-AR- 15018 and HST-GO-16196.003-A). 

\section*{Data Availability}

The data underlying this article will be shared on reasonable request to the corresponding author.



\bibliographystyle{mnras}
\bibliography{WD_bibl} 

\begin{thebibliography}{}
\makeatletter
\relax
\def\mn@urlcharsother{\let\do\@makeother \do\$\do\&\do\#\do\^\do\_\do\%\do\~}
\def\mn@doi{\begingroup\mn@urlcharsother \@ifnextchar [ {\mn@doi@}
  {\mn@doi@[]}}
\def\mn@doi@[#1]#2{\def\@tempa{#1}\ifx\@tempa\@empty \href
  {http://dx.doi.org/#2} {doi:#2}\else \href {http://dx.doi.org/#2} {#1}\fi
  \endgroup}
\def\mn@eprint#1#2{\mn@eprint@#1:#2::\@nil}
\def\mn@eprint@arXiv#1{\href {http://arxiv.org/abs/#1} {{\tt arXiv:#1}}}
\def\mn@eprint@dblp#1{\href {http://dblp.uni-trier.de/rec/bibtex/#1.xml}
  {dblp:#1}}
\def\mn@eprint@#1:#2:#3:#4\@nil{\def\@tempa {#1}\def\@tempb {#2}\def\@tempc
  {#3}\ifx \@tempc \@empty \let \@tempc \@tempb \let \@tempb \@tempa \fi \ifx
  \@tempb \@empty \def\@tempb {arXiv}\fi \@ifundefined
  {mn@eprint@\@tempb}{\@tempb:\@tempc}{\expandafter \expandafter \csname
  mn@eprint@\@tempb\endcsname \expandafter{\@tempc}}}

\bibitem[\protect\citeauthoryear{{Acero}, {Ballet}  \& {Decourchelle}}{{Acero}
  et~al.}{2007}]{Acero2007}
{Acero} F.,  {Ballet} J.,   {Decourchelle} A.,  2007, \mn@doi [\aap]
  {10.1051/0004-6361:20077742}, \href
  {https://ui.adsabs.harvard.edu/abs/2007A&A...475..883A} {475, 883}

\bibitem[\protect\citeauthoryear{{Akras}, {Guzman-Ramirez}, {Leal-Ferreira}  \&
  {Ramos-Larios}}{{Akras} et~al.}{2019}]{Akras2019}
{Akras} S.,  {Guzman-Ramirez} L.,  {Leal-Ferreira} M.~L.,   {Ramos-Larios} G.,
  2019, \mn@doi [\apjs] {10.3847/1538-4365/aaf88c}, \href
  {https://ui.adsabs.harvard.edu/abs/2019ApJS..240...21A} {240, 21}

\bibitem[\protect\citeauthoryear{{Allende Prieto}, {Lambert}  \&
  {Asplund}}{{Allende Prieto} et~al.}{2001}]{AllendePrieto}
{Allende Prieto} C.,  {Lambert} D.~L.,   {Asplund} M.,  2001, \mn@doi [\apjl]
  {10.1086/322874}, \href
  {https://ui.adsabs.harvard.edu/abs/2001ApJ...556L..63A} {556, L63}

\bibitem[\protect\citeauthoryear{{Badenes}, {Borkowski}, {Hughes}, {Hwang}  \&
  {Bravo}}{{Badenes} et~al.}{2006}]{2006Badenes}
{Badenes} C.,  {Borkowski} K.~J.,  {Hughes} J.~P.,  {Hwang} U.,   {Bravo} E.,
  2006, \mn@doi [\apj] {10.1086/504399}, \href
  {https://ui.adsabs.harvard.edu/abs/2006ApJ...645.1373B} {645, 1373}

\bibitem[\protect\citeauthoryear{{Badenes}, {Hughes}, {Bravo}  \&
  {Langer}}{{Badenes} et~al.}{2007}]{Badenes2007}
{Badenes} C.,  {Hughes} J.~P.,  {Bravo} E.,   {Langer} N.,  2007, \mn@doi
  [\apj] {10.1086/518022}, \href
  {https://ui.adsabs.harvard.edu/abs/2007ApJ...662..472B} {662, 472}

\bibitem[\protect\citeauthoryear{{Baldwin}, {Phillips}  \&
  {Terlevich}}{{Baldwin} et~al.}{1981}]{1981Baldwin}
{Baldwin} J.~A.,  {Phillips} M.~M.,   {Terlevich} R.,  1981, \mn@doi [\pasp]
  {10.1086/130766}, \href
  {https://ui.adsabs.harvard.edu/abs/1981PASP...93....5B} {93, 5}

\bibitem[\protect\citeauthoryear{{Borkowski}, {Hendrick}  \&
  {Reynolds}}{{Borkowski} et~al.}{2006}]{Borkowski2006}
{Borkowski} K.~J.,  {Hendrick} S.~P.,   {Reynolds} S.~P.,  2006, \mn@doi [\apj]
  {10.1086/508335}, \href
  {https://ui.adsabs.harvard.edu/abs/2006ApJ...652.1259B} {652, 1259}

\bibitem[\protect\citeauthoryear{{Borkowski}, {Reynolds}, {Hwang}, {Green},
  {Petre}, {Krishnamurthy}  \& {Willett}}{{Borkowski}
  et~al.}{2013}]{Borkowski2013}
{Borkowski} K.~J.,  {Reynolds} S.~P.,  {Hwang} U.,  {Green} D.~A.,  {Petre} R.,
   {Krishnamurthy} K.,   {Willett} R.,  2013, \mn@doi [\apjl]
  {10.1088/2041-8205/771/1/L9}, \href
  {https://ui.adsabs.harvard.edu/abs/2013ApJ...771L...9B} {771, L9}

\bibitem[\protect\citeauthoryear{{Borkowski}, {Gwynne}, {Reynolds}, {Green},
  {Hwang}, {Petre}  \& {Willett}}{{Borkowski} et~al.}{2017}]{Borkowski2017}
{Borkowski} K.~J.,  {Gwynne} P.,  {Reynolds} S.~P.,  {Green} D.~A.,  {Hwang}
  U.,  {Petre} R.,   {Willett} R.,  2017, \mn@doi [\apjl]
  {10.3847/2041-8213/aa618c}, \href
  {https://ui.adsabs.harvard.edu/abs/2017ApJ...837L...7B} {837, L7}

\bibitem[\protect\citeauthoryear{{Broersen}, {Chiotellis}, {Vink}  \&
  {Bamba}}{{Broersen} et~al.}{2014}]{Broersen2014}
{Broersen} S.,  {Chiotellis} A.,  {Vink} J.,   {Bamba} A.,  2014, \mn@doi
  [\mnras] {10.1093/mnras/stu667}, \href
  {https://ui.adsabs.harvard.edu/abs/2014MNRAS.441.3040B} {441, 3040}

\bibitem[\protect\citeauthoryear{{Burkey}, {Reynolds}, {Borkowski}  \&
  {Blondin}}{{Burkey} et~al.}{2013}]{Burkey2013}
{Burkey} M.~T.,  {Reynolds} S.~P.,  {Borkowski} K.~J.,   {Blondin} J.~M.,
  2013, \mn@doi [\apj] {10.1088/0004-637X/764/1/63}, \href
  {https://ui.adsabs.harvard.edu/abs/2013ApJ...764...63B} {764, 63}

\bibitem[\protect\citeauthoryear{{Byler}, {Dalcanton}, {Conroy}, {Johnson},
  {Choi}, {Dotter}  \& {Rosenfield}}{{Byler}
  et~al.}{2019}]{2019AJ....158....2B}
{Byler} N.,  {Dalcanton} J.~J.,  {Conroy} C.,  {Johnson} B.~D.,  {Choi} J.,
  {Dotter} A.,   {Rosenfield} P.,  2019, \mn@doi [\aj]
  {10.3847/1538-3881/ab1b70}, \href
  {https://ui.adsabs.harvard.edu/abs/2019AJ....158....2B} {158, 2}

\bibitem[\protect\citeauthoryear{{Cassisi}, {Iben}  \&
  {Tornamb{\`e}}}{{Cassisi} et~al.}{1998}]{cassisi1998}
{Cassisi} S.,  {Iben} Icko J.,   {Tornamb{\`e}} A.,  1998, \mn@doi [\apj]
  {10.1086/305381}, \href
  {https://ui.adsabs.harvard.edu/abs/1998ApJ...496..376C} {496, 376}

\bibitem[\protect\citeauthoryear{{Chen}, {Woods}, {Yungelson}, {Gilfanov}  \&
  {Han}}{{Chen} et~al.}{2015}]{Chen2015}
{Chen} H.-L.,  {Woods} T.~E.,  {Yungelson} L.~R.,  {Gilfanov} M.,   {Han} Z.,
  2015, \mn@doi [\mnras] {10.1093/mnras/stv1865}, \href
  {https://ui.adsabs.harvard.edu/abs/2015MNRAS.453.3024C} {453, 3024}

\bibitem[\protect\citeauthoryear{{Chevalier}, {Kirshner}  \&
  {Raymond}}{{Chevalier} et~al.}{1980}]{chevalier1980}
{Chevalier} R.~A.,  {Kirshner} R.~P.,   {Raymond} J.~C.,  1980, \mn@doi [\apj]
  {10.1086/157623}, \href
  {https://ui.adsabs.harvard.edu/abs/1980ApJ...235..186C} {235, 186}

\bibitem[\protect\citeauthoryear{{Chiotellis}, {Schure}  \&
  {Vink}}{{Chiotellis} et~al.}{2012}]{chiotellis2012}
{Chiotellis} A.,  {Schure} K.~M.,   {Vink} J.,  2012, \mn@doi [\aap]
  {10.1051/0004-6361/201014754}, \href
  {https://ui.adsabs.harvard.edu/abs/2012A&A...537A.139C} {537, A139}

\bibitem[\protect\citeauthoryear{{Chiotellis}, {Kosenko}, {Schure}, {Vink}  \&
  {Kaastra}}{{Chiotellis} et~al.}{2013}]{Chiotellis2013}
{Chiotellis} A.,  {Kosenko} D.,  {Schure} K.~M.,  {Vink} J.,   {Kaastra} J.~S.,
   2013, \mn@doi [\mnras] {10.1093/mnras/stt1406}, \href
  {https://ui.adsabs.harvard.edu/abs/2013MNRAS.435.1659C} {435, 1659}

\bibitem[\protect\citeauthoryear{{Chiotellis}, {Boumis}  \&
  {Spetsieri}}{{Chiotellis} et~al.}{2020}]{2020chiotellis}
{Chiotellis} A.,  {Boumis} P.,   {Spetsieri} Z.~T.,  2020, \mn@doi [Galaxies]
  {10.3390/galaxies8020038}, \href
  {https://ui.adsabs.harvard.edu/abs/2020Galax...8...38C} {8, 38}

\bibitem[\protect\citeauthoryear{{Chiotellis}, {Boumis}  \&
  {Spetsieri}}{{Chiotellis} et~al.}{2021}]{2021chiotellis}
{Chiotellis} A.,  {Boumis} P.,   {Spetsieri} Z.~T.,  2021, \mn@doi [\mnras]
  {10.1093/mnras/staa3573}, \href
  {https://ui.adsabs.harvard.edu/abs/2021MNRAS.502..176C} {502, 176}

\bibitem[\protect\citeauthoryear{{Choudhury}, {Subramaniam}, {Cole}  \&
  {Sohn}}{{Choudhury} et~al.}{2018}]{choud2018}
{Choudhury} S.,  {Subramaniam} A.,  {Cole} A.~A.,   {Sohn} Y.~J.,  2018,
  \mn@doi [\mnras] {10.1093/mnras/sty087}, \href
  {https://ui.adsabs.harvard.edu/abs/2018MNRAS.475.4279C} {475, 4279}

\bibitem[\protect\citeauthoryear{{Cox}}{{Cox}}{2005}]{Cox2005}
{Cox} D.~P.,  2005, \mn@doi [\araa] {10.1146/annurev.astro.43.072103.150615},
  \href {https://ui.adsabs.harvard.edu/abs/2005ARA&A..43..337C} {43, 337}

\bibitem[\protect\citeauthoryear{{Di Stefano}, {Voss}  \& {Claeys}}{{Di
  Stefano} et~al.}{2011}]{diistefano2011}
{Di Stefano} R.,  {Voss} R.,   {Claeys} J.~S.~W.,  2011, \mn@doi [\apjl]
  {10.1088/2041-8205/738/1/L1}, \href
  {https://ui.adsabs.harvard.edu/abs/2011ApJ...738L...1D} {738, L1}

\bibitem[\protect\citeauthoryear{{Dwarkadas} \& {Chevalier}}{{Dwarkadas} \&
  {Chevalier}}{1998}]{Dwarkadas1998}
{Dwarkadas} V.~V.,  {Chevalier} R.~A.,  1998, \mn@doi [\apj] {10.1086/305478},
  \href {https://ui.adsabs.harvard.edu/abs/1998ApJ...497..807D} {497, 807}

\bibitem[\protect\citeauthoryear{{Farias}, {Clocchiatti}, {Woods}  \&
  {Rest}}{{Farias} et~al.}{2020}]{Farias2020}
{Farias} D.~A.,  {Clocchiatti} A.,  {Woods} T.~E.,   {Rest} A.,  2020, \mn@doi
  [\mnras] {10.1093/mnras/staa2213}, \href
  {https://ui.adsabs.harvard.edu/abs/2020MNRAS.497.3234F} {497, 3234}

\bibitem[\protect\citeauthoryear{{Ferland} \& {Netzer}}{{Ferland} \&
  {Netzer}}{1983}]{1983ApJ...264..105F}
{Ferland} G.~J.,  {Netzer} H.,  1983, \mn@doi [\apj] {10.1086/160577}, \href
  {https://ui.adsabs.harvard.edu/abs/1983ApJ...264..105F} {264, 105}

\bibitem[\protect\citeauthoryear{{Ferland} et~al.,}{{Ferland}
  et~al.}{2017}]{Ferland2017}
{Ferland} G.~J.,  et~al., 2017, \rmxaa, \href
  {https://ui.adsabs.harvard.edu/abs/2017RMxAA..53..385F} {53, 385}

\bibitem[\protect\citeauthoryear{{Fink}, {R{\"o}pke}, {Hillebrandt},
  {Seitenzahl}, {Sim}  \& {Kromer}}{{Fink} et~al.}{2010}]{Fink2010}
{Fink} M.,  {R{\"o}pke} F.~K.,  {Hillebrandt} W.,  {Seitenzahl} I.~R.,  {Sim}
  S.~A.,   {Kromer} M.,  2010, \mn@doi [\aap] {10.1051/0004-6361/200913892},
  \href {https://ui.adsabs.harvard.edu/abs/2010A&A...514A..53F} {514, A53}

\bibitem[\protect\citeauthoryear{{Gallagher} \& {Starrfield}}{{Gallagher} \&
  {Starrfield}}{1978}]{Gallagher}
{Gallagher} J.~S.,  {Starrfield} S.,  1978, \mn@doi [\araa]
  {10.1146/annurev.aa.16.090178.001131}, \href
  {https://ui.adsabs.harvard.edu/abs/1978ARA&A..16..171G} {16, 171}

\bibitem[\protect\citeauthoryear{{Ghavamian}, {Raymond}, {Smith}  \&
  {Hartigan}}{{Ghavamian} et~al.}{2001}]{Ghavamian2001}
{Ghavamian} P.,  {Raymond} J.,  {Smith} R.~C.,   {Hartigan} P.,  2001, \mn@doi
  [\apj] {10.1086/318408}, \href
  {https://ui.adsabs.harvard.edu/abs/2001ApJ...547..995G} {547, 995}

\bibitem[\protect\citeauthoryear{{Ghavamian}, {Winkler}, {Raymond}  \&
  {Long}}{{Ghavamian} et~al.}{2002}]{Ghahavian2002}
{Ghavamian} P.,  {Winkler} P.~F.,  {Raymond} J.~C.,   {Long} K.~S.,  2002,
  \mn@doi [\apj] {10.1086/340437}, \href
  {https://ui.adsabs.harvard.edu/abs/2002ApJ...572..888G} {572, 888}

\bibitem[\protect\citeauthoryear{{Ghavamian}, {Rakowski}, {Hughes}  \&
  {Williams}}{{Ghavamian} et~al.}{2003}]{2003ghavamian}
{Ghavamian} P.,  {Rakowski} C.~E.,  {Hughes} J.~P.,   {Williams} T.~B.,  2003,
  \mn@doi [\apj] {10.1086/375161}, \href
  {https://ui.adsabs.harvard.edu/abs/2003ApJ...590..833G} {590, 833}

\bibitem[\protect\citeauthoryear{{Ghavamian}, {Blair}, {Sankrit}, {Raymond}  \&
  {Hughes}}{{Ghavamian} et~al.}{2007}]{Ghavamian2007}
{Ghavamian} P.,  {Blair} W.~P.,  {Sankrit} R.,  {Raymond} J.~C.,   {Hughes}
  J.~P.,  2007, \mn@doi [\apj] {10.1086/518686}, \href
  {https://ui.adsabs.harvard.edu/abs/2007ApJ...664..304G} {664, 304}

\bibitem[\protect\citeauthoryear{{Gomes} et~al.,}{{Gomes}
  et~al.}{2016}]{2016A&A...588A..68G}
{Gomes} J.~M.,  et~al., 2016, \mn@doi [\aap] {10.1051/0004-6361/201525976},
  \href {https://ui.adsabs.harvard.edu/abs/2016A&A...588A..68G} {588, A68}

\bibitem[\protect\citeauthoryear{{Graur} \& {Woods}}{{Graur} \&
  {Woods}}{2019}]{2019Graur}
{Graur} O.,  {Woods} T.~E.,  2019, \mn@doi [\mnras] {10.1093/mnrasl/slz005},
  \href {https://ui.adsabs.harvard.edu/abs/2019MNRAS.484L..79G} {484, L79}

\bibitem[\protect\citeauthoryear{{Grevesse} \& {Sauval}}{{Grevesse} \&
  {Sauval}}{1998}]{Grevesse1998}
{Grevesse} N.,  {Sauval} A.~J.,  1998, \mn@doi [\ssr]
  {10.1023/A:1005161325181}, \href
  {https://ui.adsabs.harvard.edu/abs/1998SSRv...85..161G} {85, 161}

\bibitem[\protect\citeauthoryear{{Gruyters}, {Exter}, {Roberts}  \&
  {Rappaport}}{{Gruyters} et~al.}{2012}]{Gruyters2012}
{Gruyters} P.,  {Exter} K.,  {Roberts} T.~P.,   {Rappaport} S.,  2012, \mn@doi
  [\aap] {10.1051/0004-6361/201219051}, \href
  {https://ui.adsabs.harvard.edu/abs/2012A&A...544A..86G} {544, A86}

\bibitem[\protect\citeauthoryear{{Hachisu}, {Kato}, {Nomoto}  \&
  {Umeda}}{{Hachisu} et~al.}{1999}]{Hachisu1999}
{Hachisu} I.,  {Kato} M.,  {Nomoto} K.,   {Umeda} H.,  1999, \mn@doi [\apj]
  {10.1086/307370}, \href
  {https://ui.adsabs.harvard.edu/abs/1999ApJ...519..314H} {519, 314}

\bibitem[\protect\citeauthoryear{{Haffner} et~al.,}{{Haffner}
  et~al.}{2009}]{Haffner2009}
{Haffner} L.~M.,  et~al., 2009, \mn@doi [Reviews of Modern Physics]
  {10.1103/RevModPhys.81.969}, \href
  {https://ui.adsabs.harvard.edu/abs/2009RvMP...81..969H} {81, 969}

\bibitem[\protect\citeauthoryear{{Heng}}{{Heng}}{2010}]{Heng2010}
{Heng} K.,  2010, \mn@doi [\pasa] {10.1071/AS09057}, \href
  {https://ui.adsabs.harvard.edu/abs/2010PASA...27...23H} {27, 23}

\bibitem[\protect\citeauthoryear{{Hillebrandt}, {Kromer}, {R{\"o}pke}  \&
  {Ruiter}}{{Hillebrandt} et~al.}{2013}]{Hillebrandt2013}
{Hillebrandt} W.,  {Kromer} M.,  {R{\"o}pke} F.~K.,   {Ruiter} A.~J.,  2013,
  \mn@doi [Frontiers of Physics] {10.1007/s11467-013-0303-2}, \href
  {https://ui.adsabs.harvard.edu/abs/2013FrPhy...8..116H} {8, 116}

\bibitem[\protect\citeauthoryear{{Holberg}}{{Holberg}}{2009}]{Holberg2009}
{Holberg} J.~B.,  2009, in Journal of Physics Conference Series. p. 012022,
  \mn@doi{10.1088/1742-6596/172/1/012022}

\bibitem[\protect\citeauthoryear{{Holweger}}{{Holweger}}{2001}]{holweger2001}
{Holweger} H.,  2001, in {Wimmer-Schweingruber} R.~F.,  ed.,  American
  Institute of Physics Conference Series Vol. 598, Joint SOHO/ACE workshop
  ``Solar and Galactic Composition''. pp 23--30 (\mn@eprint {arXiv}
  {astro-ph/0107426}), \mn@doi{10.1063/1.1433974}

\bibitem[\protect\citeauthoryear{{Iben} \& {Tutukov}}{{Iben} \&
  {Tutukov}}{1984}]{Iben1984}
{Iben} I. J.,  {Tutukov} A.~V.,  1984, \mn@doi [\apjs] {10.1086/190932}, \href
  {https://ui.adsabs.harvard.edu/abs/1984ApJS...54..335I} {54, 335}

\bibitem[\protect\citeauthoryear{{Kahabka} \& {van den Heuvel}}{{Kahabka} \&
  {van den Heuvel}}{1997}]{Kahabka1997}
{Kahabka} P.,  {van den Heuvel} E.~P.~J.,  1997, \mn@doi [\araa]
  {10.1146/annurev.astro.35.1.69}, \href
  {https://ui.adsabs.harvard.edu/abs/1997ARA&A..35...69K} {35, 69}

\bibitem[\protect\citeauthoryear{{Kato}, {Kanatsu}, {Takamizawa}, {Takao}  \&
  {Stubbings}}{{Kato} et~al.}{2000}]{Kato}
{Kato} T.,  {Kanatsu} K.,  {Takamizawa} K.,  {Takao} A.,   {Stubbings} R.,
  2000, \iaucirc, \href {https://ui.adsabs.harvard.edu/abs/2000IAUC.7552....1K}
  {7552, 1}

\bibitem[\protect\citeauthoryear{{Katsuda}, {Petre}, {Hughes}, {Hwang},
  {Yamaguchi}, {Hayato}, {Mori}  \& {Tsunemi}}{{Katsuda}
  et~al.}{2010}]{Katsuda2010}
{Katsuda} S.,  {Petre} R.,  {Hughes} J.~P.,  {Hwang} U.,  {Yamaguchi} H.,
  {Hayato} A.,  {Mori} K.,   {Tsunemi} H.,  2010, \mn@doi [\apj]
  {10.1088/0004-637X/709/2/1387}, \href
  {https://ui.adsabs.harvard.edu/abs/2010ApJ...709.1387K} {709, 1387}

\bibitem[\protect\citeauthoryear{{Kauffmann} et~al.,}{{Kauffmann}
  et~al.}{2003}]{2003MNRAS.346.1055K}
{Kauffmann} G.,  et~al., 2003, \mn@doi [\mnras]
  {10.1111/j.1365-2966.2003.07154.x}, \href
  {https://ui.adsabs.harvard.edu/abs/2003MNRAS.346.1055K} {346, 1055}

\bibitem[\protect\citeauthoryear{{Kehrig} et~al.,}{{Kehrig}
  et~al.}{2011}]{2011kehrig}
{Kehrig} C.,  et~al., 2011, \mn@doi [\aap] {10.1051/0004-6361/201015493}, \href
  {https://ui.adsabs.harvard.edu/abs/2011A&A...526A.128K} {526, A128}

\bibitem[\protect\citeauthoryear{{Kemp}, {Karakas}, {Casey}, {Izzard},
  {Ruiter}, {Agrawal}, {Broekgaarden}  \& {Temmink}}{{Kemp}
  et~al.}{2021}]{2021kemp}
{Kemp} A.~J.,  {Karakas} A.~I.,  {Casey} A.~R.,  {Izzard} R.~G.,  {Ruiter}
  A.~J.,  {Agrawal} P.,  {Broekgaarden} F.~S.,   {Temmink} K.~D.,  2021,
  \mn@doi [\mnras] {10.1093/mnras/stab1160}, \href
  {https://ui.adsabs.harvard.edu/abs/2021MNRAS.tmp.1151K} {}

\bibitem[\protect\citeauthoryear{{Kewley}, {Dopita}, {Sutherland}, {Heisler}
  \& {Trevena}}{{Kewley} et~al.}{2001}]{2001ApJ...556..121K}
{Kewley} L.~J.,  {Dopita} M.~A.,  {Sutherland} R.~S.,  {Heisler} C.~A.,
  {Trevena} J.,  2001, \mn@doi [\apj] {10.1086/321545}, \href
  {https://ui.adsabs.harvard.edu/abs/2001ApJ...556..121K} {556, 121}

\bibitem[\protect\citeauthoryear{{Kewley}, {Groves}, {Kauffmann}  \&
  {Heckman}}{{Kewley} et~al.}{2006}]{2006MNRAS.372..961K}
{Kewley} L.~J.,  {Groves} B.,  {Kauffmann} G.,   {Heckman} T.,  2006, \mn@doi
  [\mnras] {10.1111/j.1365-2966.2006.10859.x}, \href
  {https://ui.adsabs.harvard.edu/abs/2006MNRAS.372..961K} {372, 961}

\bibitem[\protect\citeauthoryear{{Kosenko}, {Vink}, {Blinnikov}  \&
  {Rasmussen}}{{Kosenko} et~al.}{2008}]{2008konsenko}
{Kosenko} D.,  {Vink} J.,  {Blinnikov} S.,   {Rasmussen} A.,  2008, \mn@doi
  [\aap] {10.1051/0004-6361:200809495}, \href
  {https://ui.adsabs.harvard.edu/abs/2008A&A...490..223K} {490, 223}

\bibitem[\protect\citeauthoryear{{Kosenko}, {Ferrand}  \&
  {Decourchelle}}{{Kosenko} et~al.}{2014}]{2014kosenko}
{Kosenko} D.,  {Ferrand} G.,   {Decourchelle} A.,  2014, \mn@doi [\mnras]
  {10.1093/mnras/stu1251}, \href
  {https://ui.adsabs.harvard.edu/abs/2014MNRAS.443.1390K} {443, 1390}

\bibitem[\protect\citeauthoryear{{Kumari}, {Maiolino}, {Belfiore}  \&
  {Curti}}{{Kumari} et~al.}{2019}]{Kumari2019}
{Kumari} N.,  {Maiolino} R.,  {Belfiore} F.,   {Curti} M.,  2019, \mn@doi
  [\mnras] {10.1093/mnras/stz366}, \href
  {https://ui.adsabs.harvard.edu/abs/2019MNRAS.485..367K} {485, 367}

\bibitem[\protect\citeauthoryear{{Kuuttila}, {Gilfanov}, {Seitenzahl}, {Woods}
  \& {Vogt}}{{Kuuttila} et~al.}{2019}]{kuuttila2019}
{Kuuttila} J.,  {Gilfanov} M.,  {Seitenzahl} I.~R.,  {Woods} T.~E.,   {Vogt}
  F.~P.~A.,  2019, \mn@doi [\mnras] {10.1093/mnras/stz065}, \href
  {https://ui.adsabs.harvard.edu/abs/2019MNRAS.484.1317K} {484, 1317}

\bibitem[\protect\citeauthoryear{{Lewis}, {Burrows}, {Hughes}, {Slane},
  {Garmire}  \& {Nousek}}{{Lewis} et~al.}{2003}]{Lewis2003}
{Lewis} K.~T.,  {Burrows} D.~N.,  {Hughes} J.~P.,  {Slane} P.~O.,  {Garmire}
  G.~P.,   {Nousek} J.~A.,  2003, \mn@doi [\apj] {10.1086/344717}, \href
  {https://ui.adsabs.harvard.edu/abs/2003ApJ...582..770L} {582, 770}

\bibitem[\protect\citeauthoryear{{Li} \& {van den Heuvel}}{{Li} \& {van den
  Heuvel}}{1997}]{Li1997}
{Li} X.~D.,  {van den Heuvel} E.~P.~J.,  1997, \aap, \href
  {https://ui.adsabs.harvard.edu/abs/1997A&A...322L...9L} {322, L9}

\bibitem[\protect\citeauthoryear{{Li} et~al.,}{{Li} et~al.}{2017}]{Li2017}
{Li} C.-J.,  et~al., 2017, \mn@doi [\apj] {10.3847/1538-4357/836/1/85}, \href
  {https://ui.adsabs.harvard.edu/abs/2017ApJ...836...85L} {836, 85}

\bibitem[\protect\citeauthoryear{{Li}, {Chu}, {Raymond}, {Leibundgut},
  {Seitenzahl}  \& {Morlino}}{{Li} et~al.}{2021}]{LI2021}
{Li} C.-J.,  {Chu} Y.-H.,  {Raymond} J.~C.,  {Leibundgut} B.,  {Seitenzahl}
  I.~R.,   {Morlino} G.,  2021, arXiv e-prints, \href
  {https://ui.adsabs.harvard.edu/abs/2021arXiv211009250L} {p. arXiv:2110.09250}

\bibitem[\protect\citeauthoryear{{Livio} \& {Mazzali}}{{Livio} \&
  {Mazzali}}{2018}]{2018Livio}
{Livio} M.,  {Mazzali} P.,  2018, \mn@doi [\physrep]
  {10.1016/j.physrep.2018.02.002}, \href
  {https://ui.adsabs.harvard.edu/abs/2018PhR...736....1L} {736, 1}

\bibitem[\protect\citeauthoryear{{Livne}}{{Livne}}{1990}]{Livne1990}
{Livne} E.,  1990, \mn@doi [\apjl] {10.1086/185721}, \href
  {https://ui.adsabs.harvard.edu/abs/1990ApJ...354L..53L} {354, L53}

\bibitem[\protect\citeauthoryear{{Lopez}, {Ramirez-Ruiz}, {Badenes},
  {Huppenkothen}, {Jeltema}  \& {Pooley}}{{Lopez} et~al.}{2009}]{Lopez2009}
{Lopez} L.~A.,  {Ramirez-Ruiz} E.,  {Badenes} C.,  {Huppenkothen} D.,
  {Jeltema} T.~E.,   {Pooley} D.~A.,  2009, \mn@doi [\apjl]
  {10.1088/0004-637X/706/1/L106}, \href
  {https://ui.adsabs.harvard.edu/abs/2009ApJ...706L.106L} {706, L106}

\bibitem[\protect\citeauthoryear{{Macchetto}, {Pastoriza}, {Caon}, {Sparks},
  {Giavalisco}, {Bender}  \& {Capaccioli}}{{Macchetto}
  et~al.}{1996}]{1996A&AS..120..463M}
{Macchetto} F.,  {Pastoriza} M.,  {Caon} N.,  {Sparks} W.~B.,  {Giavalisco} M.,
   {Bender} R.,   {Capaccioli} M.,  1996, \aaps, \href
  {https://ui.adsabs.harvard.edu/abs/1996A&AS..120..463M} {120, 463}

\bibitem[\protect\citeauthoryear{{Madej}, {Nale{\.z}yty}  \& {Althaus}}{{Madej}
  et~al.}{2004}]{Madejj2004}
{Madej} J.,  {Nale{\.z}yty} M.,   {Althaus} L.~G.,  2004, \mn@doi [\aap]
  {10.1051/0004-6361:20040120}, \href
  {https://ui.adsabs.harvard.edu/abs/2004A&A...419L...5M} {419, L5}

\bibitem[\protect\citeauthoryear{{Maoz}, {Mannucci}  \& {Nelemans}}{{Maoz}
  et~al.}{2014}]{Maoz2014}
{Maoz} D.,  {Mannucci} F.,   {Nelemans} G.,  2014, \mn@doi [\araa]
  {10.1146/annurev-astro-082812-141031}, \href
  {https://ui.adsabs.harvard.edu/abs/2014ARA&A..52..107M} {52, 107}

\bibitem[\protect\citeauthoryear{{Mart{\'\i}nez-Rodr{\'\i}guez}, {Piro},
  {Schwab}  \& {Badenes}}{{Mart{\'\i}nez-Rodr{\'\i}guez}
  et~al.}{2016}]{2016martinez}
{Mart{\'\i}nez-Rodr{\'\i}guez} H.,  {Piro} A.~L.,  {Schwab} J.,   {Badenes} C.,
   2016, \mn@doi [\apj] {10.3847/0004-637X/825/1/57}, \href
  {https://ui.adsabs.harvard.edu/abs/2016ApJ...825...57M} {825, 57}

\bibitem[\protect\citeauthoryear{{Mart{\'\i}nez-Rodr{\'\i}guez}
  et~al.,}{{Mart{\'\i}nez-Rodr{\'\i}guez} et~al.}{2018}]{martinez2018}
{Mart{\'\i}nez-Rodr{\'\i}guez} H.,  et~al., 2018, \mn@doi [\apj]
  {10.3847/1538-4357/aadaec}, \href
  {https://ui.adsabs.harvard.edu/abs/2018ApJ...865..151M} {865, 151}

\bibitem[\protect\citeauthoryear{{Miko{\l}ajewska}}{{Miko{\l}ajewska}}{2007}]{Mikolajewska2007}
{Miko{\l}ajewska} J.,  2007, Baltic Astronomy, \href
  {https://ui.adsabs.harvard.edu/abs/2007BaltA..16....1M} {16, 1}

\bibitem[\protect\citeauthoryear{{Mohamed} \& {Podsiadlowski}}{{Mohamed} \&
  {Podsiadlowski}}{2012}]{Mohamed2012}
{Mohamed} S.,  {Podsiadlowski} P.,  2012, \mn@doi [Baltic Astronomy]
  {10.1515/astro-2017-0362}, \href
  {https://ui.adsabs.harvard.edu/abs/2012BaltA..21...88M} {21, 88}

\bibitem[\protect\citeauthoryear{{Nelemans}, {Yungelson}, {Portegies Zwart}  \&
  {Verbunt}}{{Nelemans} et~al.}{2001}]{Nelemans2001}
{Nelemans} G.,  {Yungelson} L.~R.,  {Portegies Zwart} S.~F.,   {Verbunt} F.,
  2001, \mn@doi [\aap] {10.1051/0004-6361:20000147}, \href
  {https://ui.adsabs.harvard.edu/abs/2001A&A...365..491N} {365, 491}

\bibitem[\protect\citeauthoryear{{Nomoto}, {Saio}, {Kato}  \&
  {Hachisu}}{{Nomoto} et~al.}{2007}]{Nomoto2007}
{Nomoto} K.,  {Saio} H.,  {Kato} M.,   {Hachisu} I.,  2007, \mn@doi [\apj]
  {10.1086/518465}, \href
  {https://ui.adsabs.harvard.edu/abs/2007ApJ...663.1269N} {663, 1269}

\bibitem[\protect\citeauthoryear{{Osterbrock} \& {Ferland}}{{Osterbrock} \&
  {Ferland}}{2006}]{osterbrock2006}
{Osterbrock} D.~E.,  {Ferland} G.~J.,  2006, {Astrophysics of gaseous nebulae
  and active galactic nuclei}

\bibitem[\protect\citeauthoryear{{Paczy{\'n}ski}}{{Paczy{\'n}ski}}{1967}]{pazy}
{Paczy{\'n}ski} B.,  1967, \actaa, \href
  {https://ui.adsabs.harvard.edu/abs/1967AcA....17..287P} {17, 287}

\bibitem[\protect\citeauthoryear{{Pequignot}, {Petitjean}  \&
  {Boisson}}{{Pequignot} et~al.}{1991}]{Pequignot}
{Pequignot} D.,  {Petitjean} P.,   {Boisson} C.,  1991, \aap, \href
  {https://ui.adsabs.harvard.edu/abs/1991A&A...251..680P} {251, 680}

\bibitem[\protect\citeauthoryear{{Piersanti}, {Tornamb{\'e}}  \&
  {Yungelson}}{{Piersanti} et~al.}{2014}]{PIERSANTI2014}
{Piersanti} L.,  {Tornamb{\'e}} A.,   {Yungelson} L.~R.,  2014, \mn@doi
  [\mnras] {10.1093/mnras/stu1885}, \href
  {https://ui.adsabs.harvard.edu/abs/2014MNRAS.445.3239P} {445, 3239}

\bibitem[\protect\citeauthoryear{{Piersanti}, {Yungelson}  \&
  {Tornamb{\'e}}}{{Piersanti} et~al.}{2015}]{pier15}
{Piersanti} L.,  {Yungelson} L.~R.,   {Tornamb{\'e}} A.,  2015, \mn@doi
  [\mnras] {10.1093/mnras/stv1452}, \href
  {https://ui.adsabs.harvard.edu/abs/2015MNRAS.452.2897P} {452, 2897}

\bibitem[\protect\citeauthoryear{{Piersanti}, {Bravo}, {Cristallo},
  {Dom{\'\i}nguez}, {Straniero}, {Tornamb{\'e}}  \&
  {Mart{\'\i}nez-Pinedo}}{{Piersanti} et~al.}{2017}]{2017piersanti}
{Piersanti} L.,  {Bravo} E.,  {Cristallo} S.,  {Dom{\'\i}nguez} I.,
  {Straniero} O.,  {Tornamb{\'e}} A.,   {Mart{\'\i}nez-Pinedo} G.,  2017,
  \mn@doi [\apjl] {10.3847/2041-8213/aa5c7e}, \href
  {https://ui.adsabs.harvard.edu/abs/2017ApJ...836L...9P} {836, L9}

\bibitem[\protect\citeauthoryear{{Piro}, {Thompson}  \& {Kochanek}}{{Piro}
  et~al.}{2014}]{Piro2014}
{Piro} A.~L.,  {Thompson} T.~A.,   {Kochanek} C.~S.,  2014, \mn@doi [\mnras]
  {10.1093/mnras/stt2451}, \href
  {https://ui.adsabs.harvard.edu/abs/2014MNRAS.438.3456P} {438, 3456}

\bibitem[\protect\citeauthoryear{{Rakowski}, {Ghavamian}  \&
  {Hughes}}{{Rakowski} et~al.}{2003}]{2003Rakowski}
{Rakowski} C.~E.,  {Ghavamian} P.,   {Hughes} J.~P.,  2003, \mn@doi [\apj]
  {10.1086/375162}, \href
  {https://ui.adsabs.harvard.edu/abs/2003ApJ...590..846R} {590, 846}

\bibitem[\protect\citeauthoryear{{Rappaport}, {Di Stefano}  \&
  {Smith}}{{Rappaport} et~al.}{1994a}]{Rappaport1994a}
{Rappaport} S.,  {Di Stefano} R.,   {Smith} J.~D.,  1994a, \mn@doi [\apj]
  {10.1086/174106}, \href
  {https://ui.adsabs.harvard.edu/abs/1994ApJ...426..692R} {426, 692}

\bibitem[\protect\citeauthoryear{{Rappaport}, {Chiang}, {Kallman}  \&
  {Malina}}{{Rappaport} et~al.}{1994b}]{Rappaport1994}
{Rappaport} S.,  {Chiang} E.,  {Kallman} T.,   {Malina} R.,  1994b, \mn@doi
  [\apj] {10.1086/174481}, \href
  {https://ui.adsabs.harvard.edu/abs/1994ApJ...431..237R} {431, 237}

\bibitem[\protect\citeauthoryear{{Raymond}}{{Raymond}}{2001}]{Raymond2001}
{Raymond} J.~C.,  2001, \ssr, \href
  {https://ui.adsabs.harvard.edu/abs/2001SSRv...99..209R} {99, 209}

\bibitem[\protect\citeauthoryear{{Raymond}, {Korreck}, {Sedlacek}, {Blair},
  {Ghavamian}  \& {Sankrit}}{{Raymond} et~al.}{2007}]{2007Raymond}
{Raymond} J.~C.,  {Korreck} K.~E.,  {Sedlacek} Q.~C.,  {Blair} W.~P.,
  {Ghavamian} P.,   {Sankrit} R.,  2007, \mn@doi [\apj] {10.1086/512483}, \href
  {https://ui.adsabs.harvard.edu/abs/2007ApJ...659.1257R} {659, 1257}

\bibitem[\protect\citeauthoryear{{Remillard}, {Rappaport}  \&
  {Macri}}{{Remillard} et~al.}{1995}]{remillard1995}
{Remillard} R.~A.,  {Rappaport} S.,   {Macri} L.~M.,  1995, \mn@doi [\apj]
  {10.1086/175204}, \href
  {https://ui.adsabs.harvard.edu/abs/1995ApJ...439..646R} {439, 646}

\bibitem[\protect\citeauthoryear{{Rest} et~al.,}{{Rest}
  et~al.}{2005}]{Rest2005}
{Rest} A.,  et~al., 2005, \mn@doi [\nat] {10.1038/nature04365}, \href
  {https://ui.adsabs.harvard.edu/abs/2005Natur.438.1132R} {438, 1132}

\bibitem[\protect\citeauthoryear{{Reynolds}}{{Reynolds}}{1984}]{Reynolds1984}
{Reynolds} R.~J.,  1984, \mn@doi [\apj] {10.1086/162190}, \href
  {https://ui.adsabs.harvard.edu/abs/1984ApJ...282..191R} {282, 191}

\bibitem[\protect\citeauthoryear{{Ruiter}}{{Ruiter}}{2020}]{Ruiter2020}
{Ruiter} A.~J.,  2020, \mn@doi [IAU Symposium] {10.1017/S1743921320000587},
  \href {https://ui.adsabs.harvard.edu/abs/2020IAUS..357....1R} {357, 1}

\bibitem[\protect\citeauthoryear{{Ruiter}, {Belczynski}, {Sim}, {Hillebrandt},
  {Fryer}, {Fink}  \& {Kromer}}{{Ruiter} et~al.}{2011}]{Ruiter2011}
{Ruiter} A.~J.,  {Belczynski} K.,  {Sim} S.~A.,  {Hillebrandt} W.,  {Fryer}
  C.~L.,  {Fink} M.,   {Kromer} M.,  2011, \mn@doi [\mnras]
  {10.1111/j.1365-2966.2011.19276.x}, \href
  {https://ui.adsabs.harvard.edu/abs/2011MNRAS.417..408R} {417, 408}

\bibitem[\protect\citeauthoryear{{Schawinski}, {Thomas}, {Sarzi}, {Maraston},
  {Kaviraj}, {Joo}, {Yi}  \& {Silk}}{{Schawinski}
  et~al.}{2007}]{2007MNRAS.382.1415S}
{Schawinski} K.,  {Thomas} D.,  {Sarzi} M.,  {Maraston} C.,  {Kaviraj} S.,
  {Joo} S.-J.,  {Yi} S.~K.,   {Silk} J.,  2007, \mn@doi [\mnras]
  {10.1111/j.1365-2966.2007.12487.x}, \href
  {https://ui.adsabs.harvard.edu/abs/2007MNRAS.382.1415S} {382, 1415}

\bibitem[\protect\citeauthoryear{{Seitenzahl}, {Ghavamian}, {Laming}  \&
  {Vogt}}{{Seitenzahl} et~al.}{2019}]{2019Seitenzah}
{Seitenzahl} I.~R.,  {Ghavamian} P.,  {Laming} J.~M.,   {Vogt} F.~P.~A.,  2019,
  \mn@doi [\prl] {10.1103/PhysRevLett.123.041101}, \href
  {https://ui.adsabs.harvard.edu/abs/2019PhRvL.123d1101S} {123, 041101}

\bibitem[\protect\citeauthoryear{{Shen} \& {Bildsten}}{{Shen} \&
  {Bildsten}}{2007}]{shen2007}
{Shen} K.~J.,  {Bildsten} L.,  2007, \mn@doi [\apj] {10.1086/513457}, \href
  {https://ui.adsabs.harvard.edu/abs/2007ApJ...660.1444S} {660, 1444}

\bibitem[\protect\citeauthoryear{{Shen}, {Kasen}, {Miles}  \&
  {Townsley}}{{Shen} et~al.}{2018}]{Shen2018}
{Shen} K.~J.,  {Kasen} D.,  {Miles} B.~J.,   {Townsley} D.~M.,  2018, \mn@doi
  [\apj] {10.3847/1538-4357/aaa8de}, \href
  {https://ui.adsabs.harvard.edu/abs/2018ApJ...854...52S} {854, 52}

\bibitem[\protect\citeauthoryear{{Sim}, {R{\"o}pke}, {Hillebrandt}, {Kromer},
  {Pakmor}, {Fink}, {Ruiter}  \& {Seitenzahl}}{{Sim} et~al.}{2010}]{Sim2010}
{Sim} S.~A.,  {R{\"o}pke} F.~K.,  {Hillebrandt} W.,  {Kromer} M.,  {Pakmor} R.,
   {Fink} M.,  {Ruiter} A.~J.,   {Seitenzahl} I.~R.,  2010, \mn@doi [\apjl]
  {10.1088/2041-8205/714/1/L52}, \href
  {https://ui.adsabs.harvard.edu/abs/2010ApJ...714L..52S} {714, L52}

\bibitem[\protect\citeauthoryear{{Singh} et~al.,}{{Singh}
  et~al.}{2013}]{2013A&A...558A..43S}
{Singh} R.,  et~al., 2013, \mn@doi [\aap] {10.1051/0004-6361/201322062}, \href
  {https://ui.adsabs.harvard.edu/abs/2013A&A...558A..43S} {558, A43}

\bibitem[\protect\citeauthoryear{{Tian} \& {Leahy}}{{Tian} \&
  {Leahy}}{2011}]{Tian2011}
{Tian} W.~W.,  {Leahy} D.~A.,  2011, \mn@doi [\apjl]
  {10.1088/2041-8205/729/2/L15}, \href
  {https://ui.adsabs.harvard.edu/abs/2011ApJ...729L..15T} {729, L15}

\bibitem[\protect\citeauthoryear{{Toledo-Roy}, {Esquivel}, {Vel{\'a}zquez}  \&
  {Reynoso}}{{Toledo-Roy} et~al.}{2014}]{Toledo-Roy2014}
{Toledo-Roy} J.~C.,  {Esquivel} A.,  {Vel{\'a}zquez} P.~F.,   {Reynoso} E.~M.,
  2014, \mn@doi [\mnras] {10.1093/mnras/stu880}, \href
  {https://ui.adsabs.harvard.edu/abs/2014MNRAS.442..229T} {442, 229}

\bibitem[\protect\citeauthoryear{{Toonen}, {Hollands}, {G{\"a}nsicke}  \&
  {Boekholt}}{{Toonen} et~al.}{2017}]{Toonen2017}
{Toonen} S.,  {Hollands} M.,  {G{\"a}nsicke} B.~T.,   {Boekholt} T.,  2017,
  \mn@doi [\aap] {10.1051/0004-6361/201629978}, \href
  {https://ui.adsabs.harvard.edu/abs/2017A&A...602A..16T} {602, A16}

\bibitem[\protect\citeauthoryear{{Warren} \& {Hughes}}{{Warren} \&
  {Hughes}}{2004}]{warren2004}
{Warren} J.~S.,  {Hughes} J.~P.,  2004, \mn@doi [\apj] {10.1086/392528}, \href
  {https://ui.adsabs.harvard.edu/abs/2004ApJ...608..261W} {608, 261}

\bibitem[\protect\citeauthoryear{{Whelan} \& {Iben}}{{Whelan} \&
  {Iben}}{1973}]{Whelan1973}
{Whelan} J.,  {Iben} Icko J.,  1973, \mn@doi [\apj] {10.1086/152565}, \href
  {https://ui.adsabs.harvard.edu/abs/1973ApJ...186.1007W} {186, 1007}

\bibitem[\protect\citeauthoryear{{Williams} et~al.,}{{Williams}
  et~al.}{2011}]{Williams2011}
{Williams} B.~J.,  et~al., 2011, \mn@doi [\apj] {10.1088/0004-637X/741/2/96},
  \href {https://ui.adsabs.harvard.edu/abs/2011ApJ...741...96W} {741, 96}

\bibitem[\protect\citeauthoryear{{Williams}, {Borkowski}, {Ghavamian},
  {Hewitt}, {Mao}, {Petre}, {Reynolds}  \& {Blondin}}{{Williams}
  et~al.}{2013}]{2013williams}
{Williams} B.~J.,  {Borkowski} K.~J.,  {Ghavamian} P.,  {Hewitt} J.~W.,  {Mao}
  S.~A.,  {Petre} R.,  {Reynolds} S.~P.,   {Blondin} J.~M.,  2013, \mn@doi
  [\apj] {10.1088/0004-637X/770/2/129}, \href
  {https://ui.adsabs.harvard.edu/abs/2013ApJ...770..129W} {770, 129}

\bibitem[\protect\citeauthoryear{{Williams} et~al.,}{{Williams}
  et~al.}{2014}]{Williams2014}
{Williams} B.~J.,  et~al., 2014, \mn@doi [\apj] {10.1088/0004-637X/790/2/139},
  \href {https://ui.adsabs.harvard.edu/abs/2014ApJ...790..139W} {790, 139}

\bibitem[\protect\citeauthoryear{{Wolf}, {Bildsten}, {Brooks}  \&
  {Paxton}}{{Wolf} et~al.}{2013}]{wolf2013}
{Wolf} W.~M.,  {Bildsten} L.,  {Brooks} J.,   {Paxton} B.,  2013, \mn@doi
  [\apj] {10.1088/0004-637X/777/2/136}, \href
  {https://ui.adsabs.harvard.edu/abs/2013ApJ...777..136W} {777, 136}

\bibitem[\protect\citeauthoryear{{Woods} \& {Gilfanov}}{{Woods} \&
  {Gilfanov}}{2016}]{2016Woods}
{Woods} T.~E.,  {Gilfanov} M.,  2016, \mn@doi [\mnras] {10.1093/mnras/stv2423},
  \href {https://ui.adsabs.harvard.edu/abs/2016MNRAS.455.1770W} {455, 1770}

\bibitem[\protect\citeauthoryear{{Woods}, {Ghavamian}, {Badenes}  \&
  {Gilfanov}}{{Woods} et~al.}{2017}]{woods2017}
{Woods} T.~E.,  {Ghavamian} P.,  {Badenes} C.,   {Gilfanov} M.,  2017, \mn@doi
  [Nature Astronomy] {10.1038/s41550-017-0263-5}, \href
  {https://ui.adsabs.harvard.edu/abs/2017NatAs...1..800W} {1, 800}

\bibitem[\protect\citeauthoryear{{Woods}, {Ghavamian}, {Badenes}  \&
  {Gilfanov}}{{Woods} et~al.}{2018}]{Woods2018}
{Woods} T.~E.,  {Ghavamian} P.,  {Badenes} C.,   {Gilfanov} M.,  2018, \mn@doi
  [\apj] {10.3847/1538-4357/aad1ee}, \href
  {https://ui.adsabs.harvard.edu/abs/2018ApJ...863..120W} {863, 120}

\bibitem[\protect\citeauthoryear{{Woosley} \& {Kasen}}{{Woosley} \&
  {Kasen}}{2011}]{Woosley2011}
{Woosley} S.~E.,  {Kasen} D.,  2011, \mn@doi [\apj]
  {10.1088/0004-637X/734/1/38}, \href
  {https://ui.adsabs.harvard.edu/abs/2011ApJ...734...38W} {734, 38}

\bibitem[\protect\citeauthoryear{{Yamaguchi} et~al.,}{{Yamaguchi}
  et~al.}{2014}]{Yamaguchi2014}
{Yamaguchi} H.,  et~al., 2014, \mn@doi [\apjl] {10.1088/2041-8205/785/2/L27},
  \href {https://ui.adsabs.harvard.edu/abs/2014ApJ...785L..27Y} {785, L27}

\bibitem[\protect\citeauthoryear{{Yamaguchi}, {Acero}, {Li}  \&
  {Chu}}{{Yamaguchi} et~al.}{2021}]{Yamaguchi2021}
{Yamaguchi} H.,  {Acero} F.,  {Li} C.-J.,   {Chu} Y.-H.,  2021, \mn@doi [\apjl]
  {10.3847/2041-8213/abee8a}, \href
  {https://ui.adsabs.harvard.edu/abs/2021ApJ...910L..24Y} {910, L24}

\bibitem[\protect\citeauthoryear{{van den Heuvel}, {Bhattacharya}, {Nomoto}  \&
  {Rappaport}}{{van den Heuvel} et~al.}{1992}]{VandenHeuvel1992}
{van den Heuvel} E.~P.~J.,  {Bhattacharya} D.,  {Nomoto} K.,   {Rappaport}
  S.~A.,  1992, \aap, \href
  {https://ui.adsabs.harvard.edu/abs/1992A&A...262...97V} {262, 97}

\makeatother
\end{thebibliography}








\bsp	
\label{lastpage}

\appendix

\section{Modeling ionized nebulae around accreting WDs}

In this appendix, we provide for the sake of reproducibility of our results a minimal {\sc cloudy} input script, where we use to simulate the ionized nebulae around steadily nuclear burning accreting WDs. In this example, we model the nebula ionized by an one solar mass WD accreting  H-rich matter with $\Dot{M}_{\rm st}$, for ISM density 2~$\rm cm^{-3}$ and solar gas abundances. \\
\\
blackbody, T= 524807.46 K \\
luminosity linear solar 10000 \\
hden 2 linear \\
abundances "default.abn" \\
radius 16 \\
sphere \\
iterate \\
stop temperature 3e3K \\
save lines, emergent emissivity, "lines.str" last \\
O  3 5006.84A \\
O  2 3728.81A \\
O  2 3726.03A \\
O  1 6300.30A \\
H  1 4861.33A \\
H  1 6562.81A \\
N  2 6583.45A \\
He 2 4685.64A \\
end of lines \\

\end{document}